# Do wind and solar curtail at negative electricity prices?

## Incentives and evidence across two decades of German renewable support schemes

**Lion Hirth**

*Hertie School, Berlin, and Neon Neue Energieökonomik, Berlin*

hirth@neon.energy

**Abstract:** In many power systems, installed wind and solar now exceed peak electricity demand. In very sunny or windy periods, curtailing renewable generation becomes a precondition for grid stability. While turning off wind turbines and solar panels is technically easy, support schemes often give renewables an economic incentive to keep producing at negative prices. This paper studies wind and solar energy in Germany, estimating, for each cohort of generators, hour by hour the incentives implied by two decades of support policy. It then sets those incentives against observed behavior, using a new estimate of market-based curtailment built from reanalysis weather data. I find that in 2025, at prices below -50 EUR/MWh, almost all wind generators had an incentive to stop producing, but only half of them did. Nearly two thirds of solar capacity had no incentive to curtail at all, mostly because it receives a feed-in tariff that shields it from wholesale prices. Of the exposed remainder, about one in seven cut production. Low exposure and response rates inflate subsidy payments and make the power system harder to operate safely. I conclude that a further expansion of wind and solar requires them to respond to price signals, which depends on policy makers providing proper incentives and on generators acting on them.

# 1. Introduction

Wind turbines and solar panels have become so cheap that cost-optimal power system configurations often have far more wind and solar capacity than peak demand. Germany and Spain already operate wind and solar capacity of roughly twice their peak load. This implies that in very windy or sunny conditions, not all energy that could be produced is actually produced. Curtailment of excess energy is part of efficient system operation, not a failure of it. In a liberalized market the wholesale price is the main mechanism to signal scarcity and abundance: a price falling below zero signals that wind and solar generators should reduce output.

While negative wholesale prices are nothing new, they have become much more common in recent years, now making up around 6 % of all hours in many European power markets (Figure 1), and even more in some parts of Australia, the US, Chile, and other markets (AEMO, 2026; EIA, 2025; ACERA, 2025).

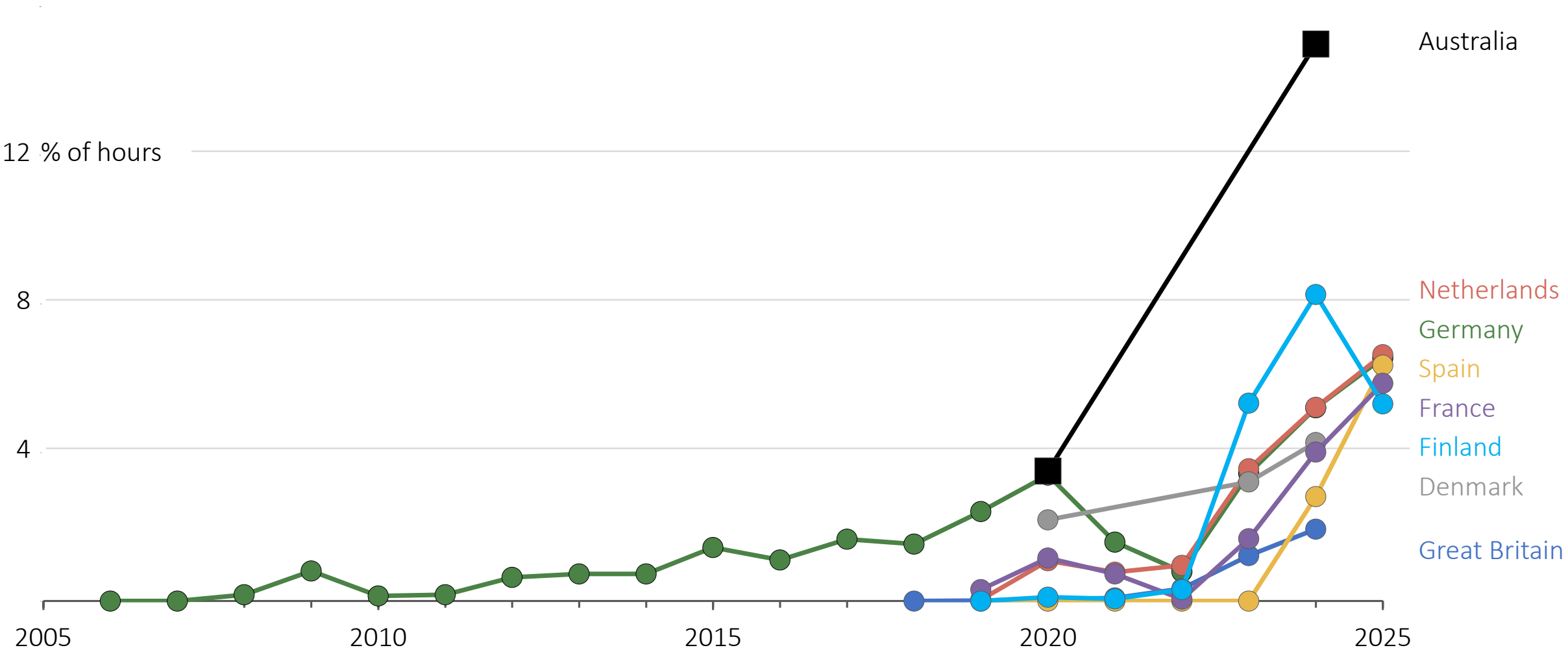

Figure 1. Share of time with negative day-ahead prices. Sources: national regulators, TSOs, market operators and ENTSO-E.

Yet some wind and solar generators keep producing at negative prices, e.g., because subsidies are paid per MWh generated. This is true for feed-in tariffs and hourly contracts for difference, which shield generators entirely from price signals, but also for green certificates, production tax credits and CfDs with reference periods, which do not block but distort the price signal. In addition, some generators that are exposed to electricity prices do not curtail, possibly because the fixed costs of dispatch processes are too high.

This paper studies market-based curtailment empirically: Do wind and solar generators stop producing when they should? The question has two parts. The first is about incentives: Which financial incentives did wind and solar generators have to curtail their output at negative electricity prices? How far have subsidies distorted the production decision? The second part is about behavior: To what degree did generators follow these incentives? I address these questions with two decades of German support policy and electricity market data.

The analysis proceeds in three steps. First, I derive the financial implications for each of the multiple support policies that Germany has implemented. For some policies, the implied incentives have to be calculated for every hour individually. Second, I estimate the market-based curtailment of wind and solar energy. This requires a new time series of the technical generation potential from historical weather data, calibrated so as to avoid endogeneity. Third, I compare expected curtailment (according to incentives) with observed curtailment to determine the share of generation that responds to incentives.

After a series of policy reforms, almost the entire German wind fleet is exposed to wholesale prices: at deeply negative prices of -50 EUR/MWh, almost all wind turbines have an incentive to curtail. However, only half the capacity actually does so. For solar, the numbers are much lower. At prices below -50 EUR/MWh, just 39 % of solar generation had an incentive to curtail, of which about one in seven did, so that only 6 % of the generation was curtailed. The remaining 94 % of the solar potential kept producing.

This is the first paper that systematically calculates financial production (or curtailment) incentives of the entire fleet of wind and solar energy of a major electricity system and sets them against observed behavior. It makes three contributions. First, from a detailed reading of two decades of regulation it derives, cohort by cohort and hour by hour, the price at which each part of the fleet should stop producing. Second, it estimates the technical production potential from reanalysis data to derive a novel time series of market-based curtailment. Third, comparing the two, it measures the degree to which generators respond to incentives.

Section 2 reviews the literature. Section 3 derives the incentives that different support schemes create. Section 4 quantifies them for Germany. Section 5 estimates market-based curtailment from weather data. Section 6 compares incentives and behavior. Section 7 discusses implications and Section 8 concludes.

# 2. Related literature

The literature relevant to this paper falls into four branches. The first is theoretical and derives how support instruments distort the production decision. The second measures curtailment volumes and is concerned mainly with grid-related curtailment. The third estimates empirically how wind and solar generators respond to negative prices under specific schemes; it is closest to this paper. The fourth consists of market analyses that estimate market-based curtailment volumes.

Several dozen papers discuss how the design of a support instrument shapes the incentive to produce at low or negative prices. Klessmann et al. (2008) set out the trade-off between exposing renewables to market risk and preserving their dispatch incentives. Kitzing (2014) formalizes the risk side of feed-in tariffs against premiums. Andor and Voss (2016) show that generation-based subsidies distort dispatch where the externality is not proportional to output. Pahle et al. (2016) model the welfare effects of premiums that induce generation at negative prices. Bunn and Yusupov (2015) and Newbery (2023) analyze why conventional contracts for difference blunt the price signal and how a reference-volume design restores it. Schlecht et al. (2024) develop the argument for financial contracts. This branch establishes the mechanisms used in Section 3.

The second branch measures curtailment volumes. Joos and Staffell (2018) report wind curtailment of 4 to 5 % of output in Germany and 5 to 6 % in Great Britain for 2010 to 2016, from Balancing Mechanism instructions in Britain, which cover transmission-connected farms only, and from the German compensation statistics for feed-in management. Bird et al. (2016) compile operator-reported volumes for eleven countries and note that most do not separate causes. The curtailment-energy share of Yasuda et al. (2022) uses the same statistics and covers grid-related curtailment only, except for Denmark and the United States, where the statistics do not separate market-based curtailment. These data therefore record almost only grid-related curtailment. López Prol and Zilberman (2023) estimate economic curtailment econometrically from hourly CAISO data. None of these papers measures market-based curtailment in a European zonal electricity market.

A third, smaller branch estimates how far generators respond to negative prices under a given support scheme. This is the question of the present paper. Purkus et al. (2015) combine interviews with direct marketers and agent-based simulation of the German market premium in its first years. They find that wind plants in direct marketing stopped producing at around -65 EUR/MWh, that is, at minus the premium then in force, and that the scheme changed little else in dispatch. The finding is based on stated behavior and a simulation model and does not use observed output. Frondel et al. (2022) estimate a counterfactual with Bayesian additive regression trees on hourly prices from 2009 to 2016. They attribute to the 2012 market premium a reduction of about 70 % in the number of negative-price hours. The study covers the whole market but observes prices only. It cannot tell which plants responded or how much energy they curtailed.

Van Steenberghe and Ovaere (forthcoming) is the study most closely related to this one. They use unit-level metered output of British offshore wind farms from 2019 to 2024 and compare, hour by hour, farms under contracts for difference with merchant farms in negative day-ahead hours. Merchant farms cut output by 69 to 83 % in those hours; CfD-backed units did so only once the six-consecutive-hour rule suspended their difference payments. In the balancing market they curtailed 28 % less when the payment covered negative imbalance prices. Their identification relies on the contrast between contract types within one technology and one market. The present paper differs in four respects. First, it covers an entire national fleet of wind and solar and not one technology segment. Second, it measures the incentive as a curtailment threshold per cohort that moves with capture prices and episode length, where they compare contract types. Third, it spans twenty years and the full sequence of German rule changes. Fourth, it estimates the generation potential from weather data instead of declared availability and nets out grid-related curtailment.

The last branch consists of consultant reports that are not peer-reviewed. Kern et al. (2024) is the only study I am aware of that quantifies the incentive side: from the plant register it finds that 75 % of solar and 67 % of wind capacity holds a fixed tariff or an unrestricted premium and therefore has little incentive to respond. Dexter Energy (2025) estimates that about 11 % of potential Dutch solar output is curtailed through imbalance-market arbitrage, Storpeak (2026) puts Spanish solar curtailment at 2 to 3 % of potential, and Montel (2026) reports 1.5 TWh of German market-based curtailment in the first half of 2026. Little is disclosed about how these numbers are produced.

Three gaps remain. First, no study compares the curtailment that a whole fleet should undertake with the curtailment it does undertake. Second, the incentive has so far been represented only by a reform date or a contract type, so the price at which each part of a fleet should stop producing is unknown.

Third, there is no long-run series of market-based curtailment, because the quantity is recorded in no statistic. This paper addresses these three gaps.

Table 1. Empirical studies of the response of wind and solar generators to negative prices, peer-reviewed and grey.

| Study | Coverage | Data and method | Main finding on (non-)curtailment |
|---|---|---|---|
| Peer-reviewed publications | | | |
| Purkus et al. (2015) | Germany, 2012 to 2014 | Interviews and agent-based modeling | Wind in direct marketing curtails at about -65 EUR/MWh, that is at minus the premium; little other price response |
| Frondel et al. (2022) | Germany, 2009 to 2016 | BART counterfactual on hourly day-ahead prices | The 2012 market premium with direct marketing cut negative-price hours by about 70 % |
| Van Steenberghe and Ovaere (forthcoming) | Great Britain, 2019 to 2024 | Unit-level offshore wind, CfD versus merchant | Merchant farms cut output 69 to 83 % in negative hours, CfD units only once the six-hour rule binds; 2.9 TWh excess generation, GBP 176 m support cost |
| Consultant reports | | | |
| Dexter Energy (2025) | Netherlands, 2024 to 2025 | Market and imbalance data | About 11 % of potential Dutch solar output curtailed through imbalance-market arbitrage |
| Storpeak (2026) | Spain, 2024 to 2025 | Market data | Solar curtailment of about 2 % of potential in 2024 and 2.9 % on average in 2024 and 2025 |
| Montel (2026) | Ten European markets, first half of 2025 and 2026 | Market data | German market-based curtailment up from 1.2 to 1.5 TWh while negative-price hours fell from 389 to 299 |

# 3. The incentive to (not) curtail

Merchant wind and solar plants produce as much as the weather allows when the wholesale price is positive. They stop once it turns negative to avoid making a loss. Plants that receive support per MWh of output behave differently. This section derives how support schemes, and some commercial contracts, distort the production decision.

## 3.1. Marginal revenue and the curtailment threshold

A firm produces as long as the price it receives for an additional unit exceeds the cost of producing it. For thermal plants that cost is dominated by fuel and emission allowances, which sets the price below which they stop producing. For wind and solar, which have no variable cost, the curtailment threshold is zero. Curtailing them is cheap and reversible, unlike a thermal shutdown. A thermal unit that stops incurs start-up fuel and thermal stress and must respect minimum stable output while online and minimum downtime once offline. Wind turbines pitch their blades out of the wind and solar inverters reduce output within seconds and at no cost.

Per-MWh subsidy payments tied to actual generation shift the curtailment threshold below zero. Such payments are an opportunity cost of curtailing: a plant that stops forgoes the payment, so the payment enters the production decision like a negative marginal cost. The plant keeps producing as long as the market price plus the payment at stake is positive. The supply curve with support is therefore a merit order of opportunity costs that extends below zero. The curtailment threshold $p^*$ is the day-ahead price below which curtailing is the profit-maximizing choice; it equals minus the support at stake per MWh:

$$p^* = -\sigma \tag{1}$$

There is no single wholesale price but a sequence of them, from the day-ahead auction through continuous intraday trading to the imbalance price. In principle the relevant price is the last one before delivery; in practice it is often the day-ahead price.

## 3.2. Subsidies distort the curtailment threshold

Under a feed-in tariff, the plant receives a fixed price $f$ per MWh fed in, irrespective of the wholesale price. Marginal revenue $MR$ equals $f$ and is positive in every hour. No price is low enough to make curtailing pay, so no curtailment threshold exists.

Under a fixed premium, a fixed amount $s$ per MWh is paid on top of market revenues, so marginal revenue is $p + s$. The premium is set by policy and does not move with market prices. The curtailment threshold is $-s$.

Under a two-sided contract for difference settled against the hourly day-ahead price, revenue per MWh equals the strike price $k$ in every hour, whatever the market price does. Marginal revenue is $k$, so the wholesale price never enters the production decision. As under a feed-in tariff, no curtailment threshold exists.

Under a two-sided contract for difference settled against a period average $r$, marginal revenue is $p + (k - r)$, with no floor at zero. When the capture price exceeds the strike price the per-MWh settlement becomes a charge. The plant then maximizes profit by curtailing even at positive prices. The German skimming of surplus revenues in 2022 and 2023 is an example: it was computed on the monthly capture price rather than on the plant's own hourly revenues. The inframarginal revenue cap of Regulation (EU) 2022/1854 works in the same way. Here the curtailment threshold is $r - k$, positive whenever the reference exceeds the strike price.

Under a one-sided contract for difference settled against the hourly day-ahead price, the premium fills the gap between the strike price $k$ and the price of the same hour, so marginal revenue is $\max\{p, k\}$. It is positive in every hour as long as the strike price is positive. The curtailment threshold is again absent.

Under a one-sided contract for difference against a period average, also called sliding premium and, in Germany, market premium, the premium fills the gap between a strike price $k$ and the average capture price $r$ of the reference period, floored at zero. Within the period it acts like a fixed premium, so marginal revenue is $p + \max\{k - r, 0\}$. Reference periods differ across schemes, from monthly to annual. The difference matters for the results (Appendix A3). Two properties are relevant empirically. First, the premium, and with it the curtailment threshold, moves with market prices and is not set by policy. Second, the premium is zero in periods in which the capture price reaches the strike price. The curtailment threshold is $-\max\{k - r, 0\}$.

Under tradable certificates the plant earns $n$ certificates per MWh, worth $c$ each, so marginal revenue is $p + n \cdot c$. Certificate values have been high relative to premiums, which implies particularly low curtailment thresholds. The curtailment threshold is $-n \cdot c$.

Support that is not paid per MWh generated, such as capacity-based or investment support, leaves marginal revenue at the market price (Andor and Voss, 2016). Minimum-operating-hour conditions can re-introduce a weak incentive to keep generating. The curtailment threshold stays at zero, as for a plant without support.

The instruments above, capacity-based support excepted, pay per MWh generated and therefore distort the production decision. Support can instead be made independent of the plant's own output: a fixed capacity payment, a financial contract for difference settled on a reference volume such as the output of a reference plant or a share of fleet output, or a forward contract that the generator settles financially. Marginal revenue then equals the market price in every hour, while the investor remains hedged against price risk. The open question is how much of the risk reduction is lost when the contracted volume no longer tracks the plant's own production (Schlecht et al., 2024; Sánchez Canales and Hirth, 2026). Two-sided CfDs become the European default for new plants from 2027, which makes the choice between production-based and production-independent settlement a current design question (Newbery, 2023; Schlecht et al., 2024). Germany has not used production-independent support for wind and solar. As without support, the curtailment threshold is zero.

Table 2. Typology of support instruments and implied curtailment thresholds.

| Instrument | Reference price | Marginal revenue | Curtailment threshold $p^*$ | Use in Germany |
|---|---|---|---|---|
| Feed-in tariff | - | $MR = f$ | No incentive at any price | EEG tariff: all plants to 2011, optional to 2015, below 100 kW since 2016 (rooftop solar) |
| Fixed premium per MWh | - | $MR = p + s$ | $p^* = -s$ | - |
| Two-sided CfD, hourly | Hourly day-ahead price | $MR = k$ | No incentive at any price | - |
| Two-sided CfD, reference | Period average capture price $r$ | $MR = p + (k - r)$ | $p^* = r - k$ (positive when $r > k$) | Revenue skim Dec 2022 to Jun 2023: supported plants above 1 MW, monthly; to be introduced 2027 |
| One-sided CfD, hourly | Hourly day-ahead price | $MR = \max\{p, k\}$ | No incentive at any price | - |
| One-sided CfD, reference | Period average capture price $r$ | $MR = p + \max\{k - r, 0\}$ | $p^* = -\max\{k - r, 0\}$ | EEG market premium since 2012 (monthly): wind, solar from 100 kW; mandatory from 500 kW (2014) and 100 kW (2016); auctions since 2017 |
| Tradable certificates per MWh | - | $MR = p + n \cdot c$ | $p^* = -n \cdot c$ | - |
| Capacity or investment payment | - | $MR = p$ | $p^* = 0$ | - |

Policy makers have acknowledged that per-MWh support keeps plants generating when the system does not need the energy. In response, many schemes have introduced negative-price provisions that suspend support payments when prices are negative. Sometimes such suspensions are only triggered if prices become negative for a certain amount of time (“6-hour rule”). Some schemes return the withheld payments later, for example by extending the support period.

## 3.3. Commercial contracts and self-consumption

Merchant plants and plants under financial PPAs settled against an index face the hourly price and therefore an undistorted curtailment threshold of zero. Under a pay-as-produced PPA with a fixed price per MWh delivered, the plant faces the same incentive as under a feed-in tariff: the offtaker bears the price risk. Whether the plant stops depends on who holds the dispatch right and on the contract's negative-price clause. Recent contracts floor settlement at zero or suspend delivery in negative hours, which restores a curtailment threshold of zero. How common each clause is cannot be observed, because PPA terms are not public and no register records them. For Germany this matters little: contracted PPA volumes are small compared to the supported fleet.

Guarantees of origin sell the green attribute separately from the energy and are issued per MWh, so their value enters marginal revenue exactly like a certificate and the curtailment threshold equals minus the price of the guarantee, typically a few EUR/MWh.

Most supported plants in Germany sell through an aggregator (“direct marketer”), which holds the dispatch right. Whether the price signal reaches the plant depends on the contract between owner and aggregator, in particular on whether curtailed energy is compensated and how imbalance costs are shared. As with PPAs, little is known publicly about these contracts.

For self-consumed energy the relevant price is not the wholesale price but the retail tariff that the consumer avoids. This is a major reason why solar does not respond to negative prices. Germany is the extreme case, because of the large role of behind-the-meter solar energy, and because retail contracts are almost universally fixed-price, so nothing in the consumer's bill changes when the wholesale price turns negative.

## 3.4. The merit order with curtailment

Most European countries operate multiple support schemes in parallel, for different asset sizes and historical cohorts. Conceptually, one can determine the opportunity cost to curtail – the curtailment threshold – for each MW of capacity and draw a short-term supply curve, or “merit order” curve (Figure 2).

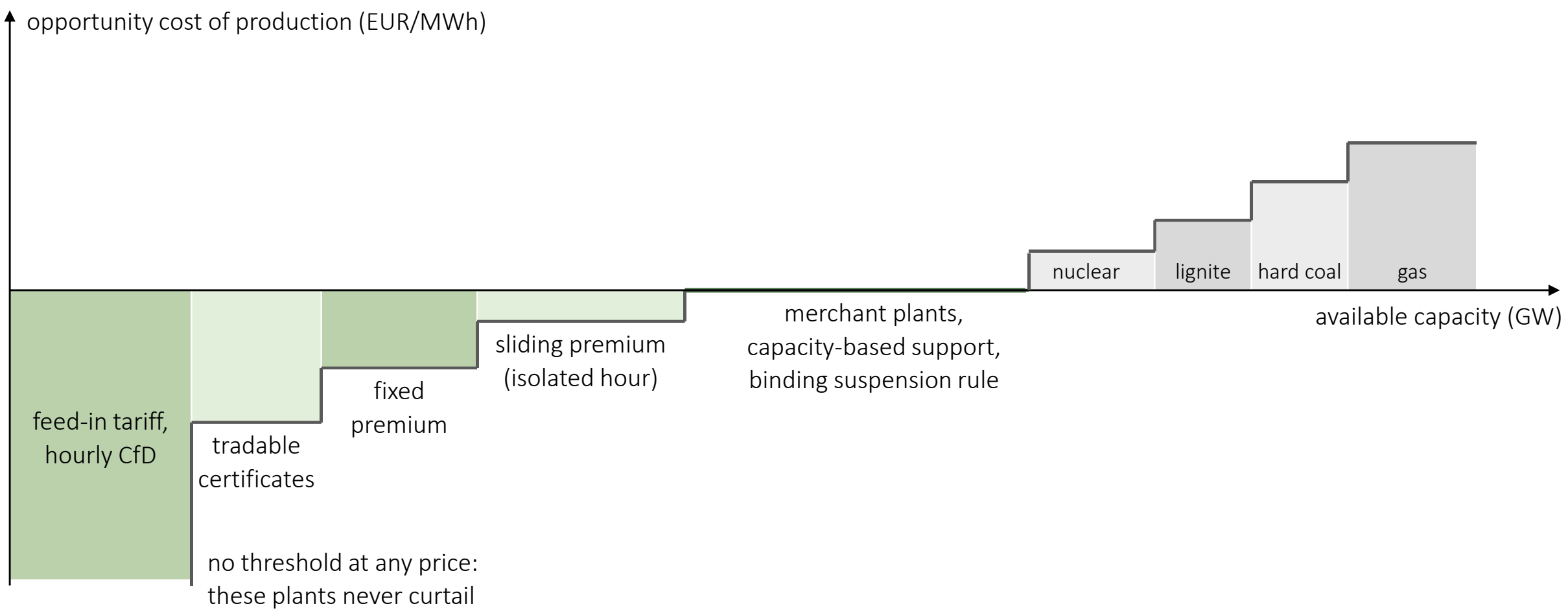

Figure 2. The merit order, accounting for support schemes (schematic).

# 4. Quantifying curtailment incentives

This section quantifies how much German capacity can be expected to curtail at which price. This is laborious, because several support schemes run in parallel, their parameters have been revised repeatedly while continuing to apply to older cohorts, and curtailment thresholds move with capture prices. Rather than calculating a single merit order curve per year, one needs to be estimated for every hour.

## 4.1. Methods and data

The first step is to cluster capacity into cohorts, that is, groups in which the same curtailment threshold applies. These differ by technology, commissioning year, support instrument and size class, which results in a total of 30 cohorts. The size of each cohort (MW) is estimated from national statistics and the public plant register. A cohort is, for instance, the 10.05 GW of onshore wind built 2016–2020 in turbines of at least 3 MW (average strike price of 76 EUR/MWh, six-hour rule applies); another is the 9.72 GW of solar assets of at least 400 kW commissioned between 2023 and 24 February 2025 (strike price 65 EUR/MWh, three-hour rule applies).

For assets under the CfD, the payment depends on the monthly (or annual) fleet capture prices. I assume that generators anticipate the capture price correctly and therefore use realized capture prices. For every cohort and month I determine the support level, the negative-price provision, and the capacity to which it applies. Appendix A2 details the rules and parameters.

Germany has witnessed more than 3,200 hours with negative day-ahead prices since the first one in October 2008. For each hour, the curtailment threshold depends on the length of the negative-price periods, because this determines if suspension rules apply. In total, I checked more than 70,000 cohort-hour combinations, of which more than 30,000 showed a negative profit margin (day-ahead price minus curtailment threshold) and thus an incentive to curtail. In the others, support payments exceeded the negative price.

Three caveats apply. First, the capacity of each cohort is my own estimate rather than statistical data. Second, Germany pays higher support for wind turbines at poorer sites, so strike prices differ within an onshore wind cohort. I use the capacity-weighted average strike price of each cohort. Splitting the cohorts by resource quality into 84 sub-cohorts changes the results little (Appendix A3). Third, for some cohorts, suspended payments are returned by extending the support period. I ignore the financial incentives this implies in calculating the profit margin.

## 4.2. Merit orders with curtailment in 2025

To see the consequences of these complicated rules, take the example of 23 June 2025 at 13:00, the hour with the highest combined wind and solar potential of the year. Figure 3 shows the German merit order curve (short-term supply curve) for this hour. For those generators settling on the monthly capture price, market premium payments are large because of the low capture prices of that month. Of the 99 GW total potential, around 30 GW had no incentive to curtail at any price, most of it rooftop solar. About 50 GW had a curtailment threshold between -120 and -1 EUR/MWh. 21 GW should curtail at zero, including merchant solar and offshore wind, but also the onshore wind under the six-hour rule, because that day prices were negative seven hours in a row. At the actual day-ahead price of -26 EUR/MWh, these calculations suggest that 50 GW had an incentive to curtail. In fact, however, only 21 GW curtailed.

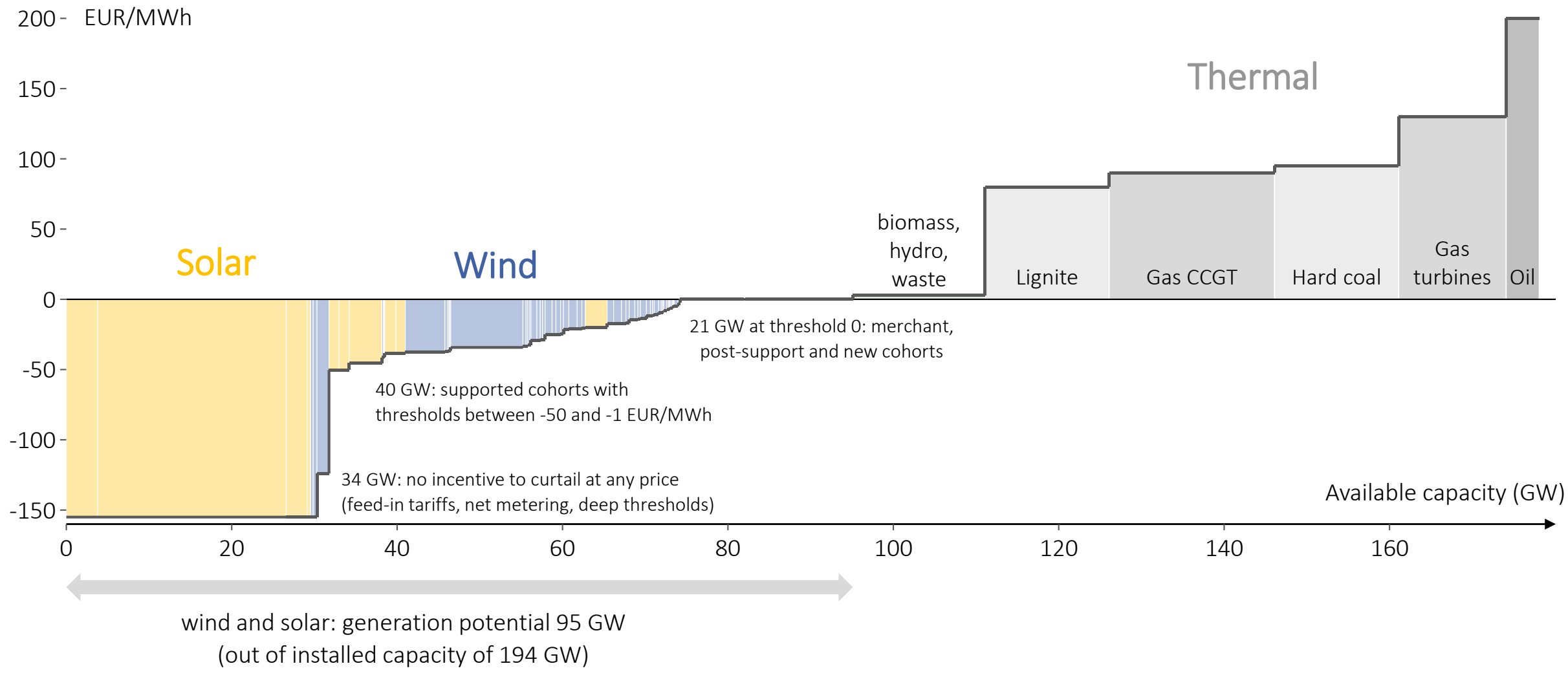

Figure 3. The German merit order with curtailment, 23 June 2025, 13:00. Opportunity costs of wind and solar are estimated from regulation; variable cost of thermal plants derived from commodity costs at that time.

The curtailment thresholds of the wind and solar fleets vary within 2025, because of the monthly capture price and the six-, four- and three-hour rules (Figure 4). The dashed curves show the effect of the suspension rules. The vast majority of the 78 GW of installed wind capacity had an incentive to curtail once prices reached -40 EUR/MWh. For solar, around 80 GW still had an incentive to produce at that price. Just above half of the solar capacity had no incentive to stop at any price.

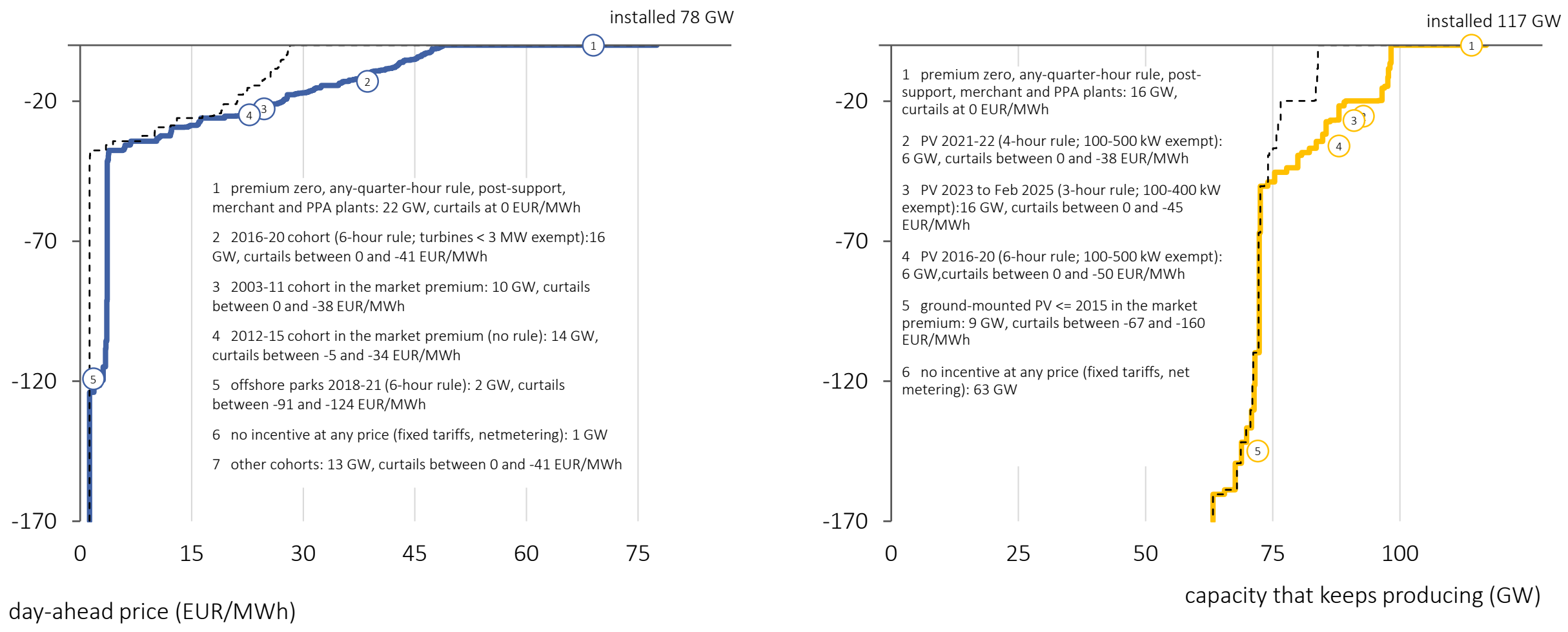

Figure 4. Curtailment thresholds of wind and solar cohorts (2025). Isolated negative hour (solid) and episode of at least six consecutive negative hours (dashed).

## 4.3. Development over time

The capacity that has an incentive to curtail at a given negative price has changed over time (Figure 5). For wind energy, this is mostly due to three independent changes: growth, policy and prices. First, more wind capacity was installed. Second, the one-sided CfD was introduced in 2012, replacing the feed-in tariff. Later, annual reference periods replaced the monthly reference. Suspension rules were tightened from six to three to two to one hour and finally to a quarter-hour. The third factor is electricity prices. With higher electricity prices during the energy crisis of 2021–22, the capture price exceeded the strike prices and support payments fell to zero, and with them their distortive effect. Solar is different, because until recently most new capacity was small-scale and continued to fall under the feed-in tariff, which provides no incentive to curtail, whatever the wholesale price.

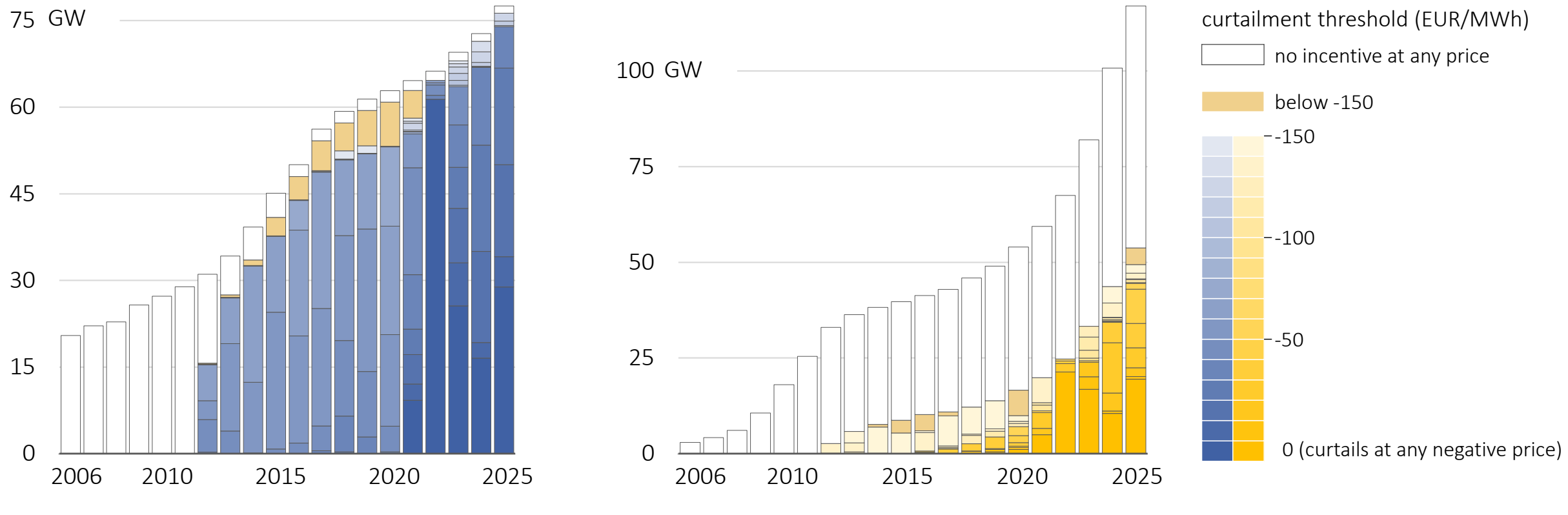

Figure 5. Wind and solar capacity by curtailment threshold for wind (left) and solar (right). Isolated negative hour, 2006 to 2025.

Another way to look at the same data is to estimate the capacity that has no incentive to curtail at any price (Figure 6). Despite policy reform efforts to turn the trend, this stock has grown every year since 2016. Most of the growth comes from rooftop solar. Since 2024 it has exceeded average load.

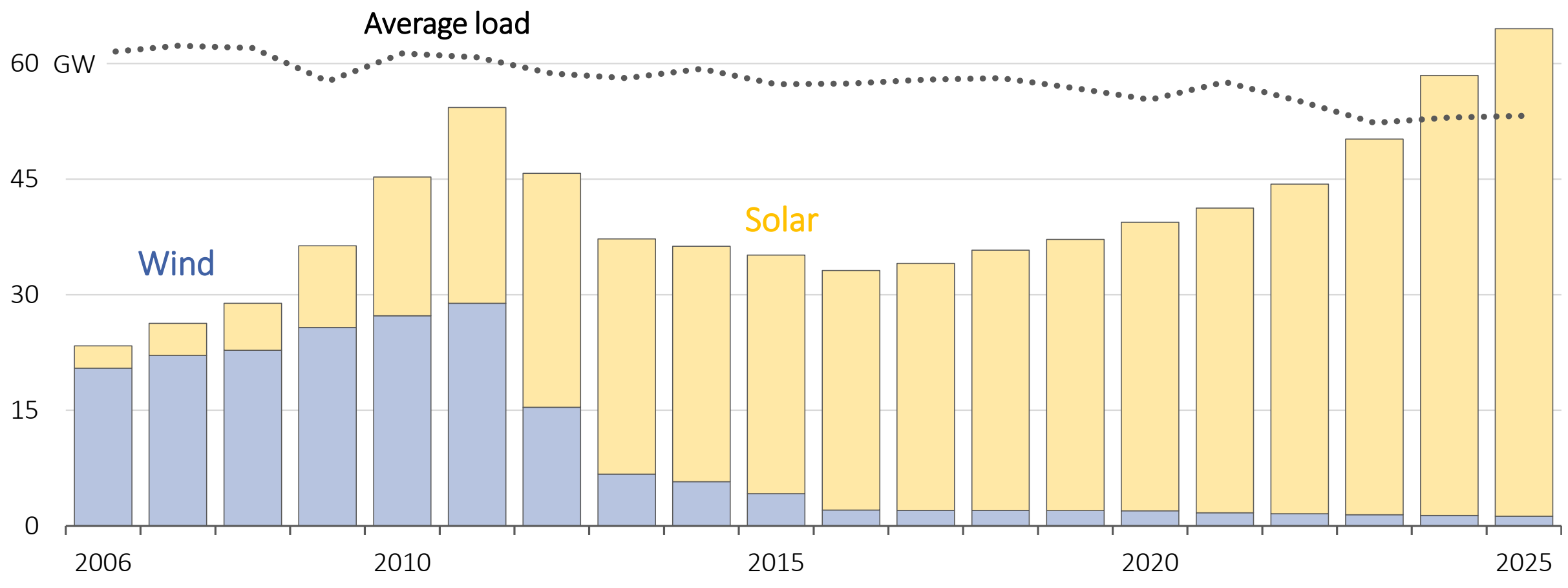

Figure 6. Installed wind and solar capacity with no incentive to curtail at any price.

# 5. Actual curtailment

The previous section determined how much capacity *should* curtail, given the incentives of the support schemes. This section asks how much generation *was* in fact curtailed. Unlike grid-related curtailment, market-based curtailment is recorded in no statistic. I therefore estimate curtailment as the gap between the generation potential and actual generation.

## 5.1. Methods and data

I estimate market-based curtailment as generation potential minus actual generation minus grid-related curtailment for every hour. The generation potential comes from a weather-based model of fleet output that is fitted to metered generation. The difficulty is that in the hours of interest metered generation is reduced by the curtailment I want to measure. A model fitted on all hours would attribute curtailment to its coefficients and find none. The estimate therefore proceeds in three steps. First, I take actual generation and add grid-related curtailment, both of which are provided by TSOs.

Second, I select the hours in which no plant had a price incentive to curtail. I exclude all hours with a day-ahead price below 5 EUR/MWh or an imbalance price below -50 EUR/MWh.

Third, I use the remaining data to calibrate a physical model of wind and solar generation based on reanalysis data. The weather input is the ERA5 reanalysis for 2001 to 2026: 100 m wind speed at seventeen grid points, twelve onshore and five offshore, and surface irradiance. A fleet-level power curve with an explicit storm-shutdown term is fitted per technology on the calibration hours. Capacity follows official year-end statistics with monthly registry additions for the within-year path. Hirth (2026) documents the methodology and its validation in more detail. Appendix A5 reports robustness checks.

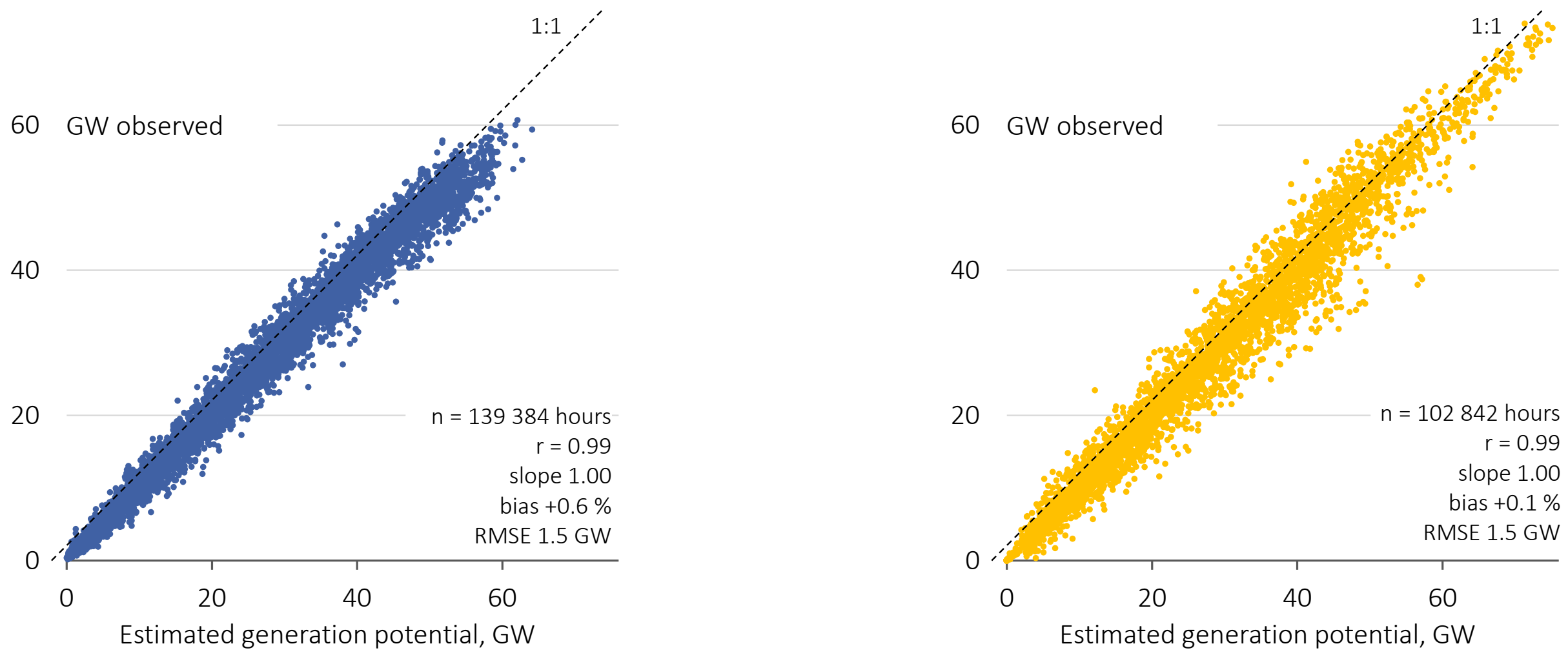

Figure 7. Generation potential against observed generation in the calibration hours for wind (left) and solar (right). 2010 to October 2026. Observed = metered generation plus grid-related curtailment. The sample is stratified over the estimate, in bins of 2.5 GW with the same quota in every bin. The statistics are computed on all calibration hours (139,384 for wind and 102,842 for solar). Dashed line: 45 degrees.

The analysis has three limitations. First, because market-based curtailment is measured as a residual, any estimation error in the generation potential is attributed to curtailment. Hirth (2026) documents the steps taken to avoid bias. Estimating generation potential from reanalysis data nevertheless remains error-prone. Second, grid-related curtailment data are coarser before October 2021, which is why I report only data from 2022 onwards in this section. Finally, "observed generation" is an estimate itself: hourly wind and solar generation is not metered plant by plant but extrapolated by the transmission system operators from a sample of plants. Any systematic bias in the negative-price hours would bias all estimates of market-based curtailment.

## 5.2. Observed curtailment

Market-based curtailment appears in significant volumes from 2022 for wind and from 2025 for solar (Figure 8). Volumes are the residual summed over the hours with a negative day-ahead price. In 2025 the volume was 2.1 TWh for wind and 0.6 TWh for solar, and 2.5 and 1.8 TWh in the first ten months of 2026, broadly consistent with Montel (2026). Grid-related curtailment remains larger, about 7 TWh for wind in 2025. The wind volumes before 2022 are upper bounds, because grid-related curtailment is available only as an annual total allocated pro rata to generation until October 2021, which understates it in the hours with negative prices.

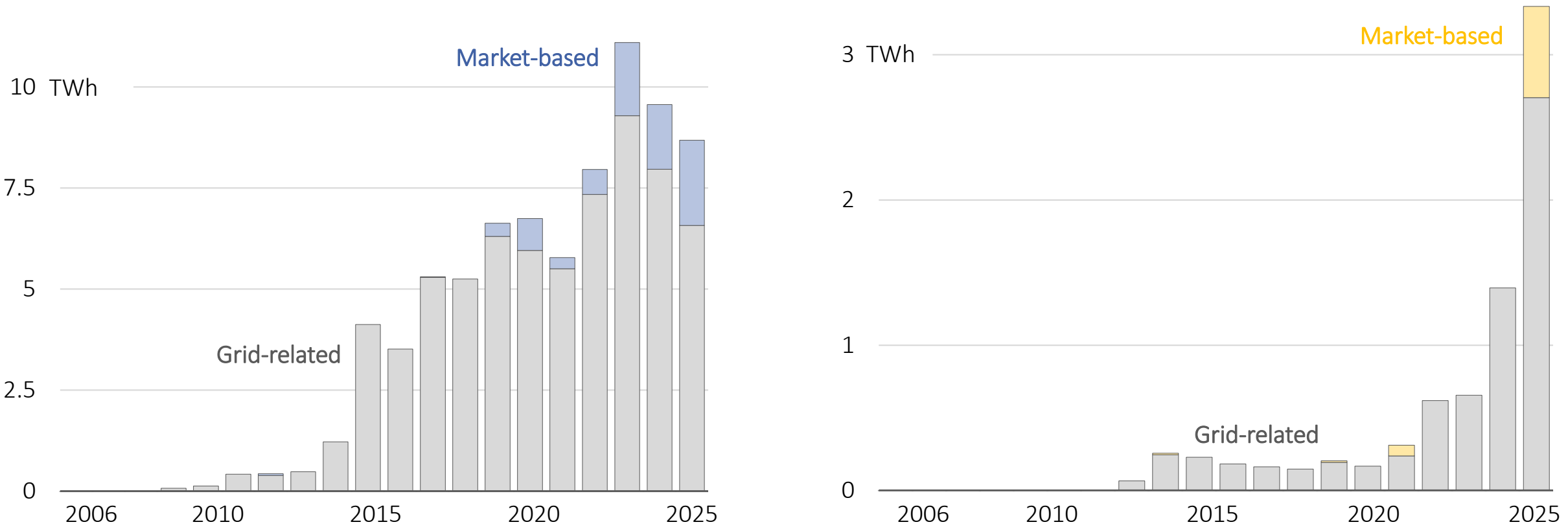

Figure 8. Market-based and grid-related curtailment for wind (left) and solar (right). Market-based curtailment is the residual summed over the hours with a negative day-ahead price; grid-related curtailment is the annual total.

The deeper the price falls, the larger the share of the generation potential that is curtailed (Table 3). Section 6 shows that two factors drive this: more generators have an incentive to curtail, and a larger fraction of those responds to the incentives. Pooled over 2022 to October 2026, 57 % of the wind potential but only 8 % of the solar potential was curtailed in the hours below -100 EUR/MWh. Both shares have grown: in 2026 they stand at 69 and 14 % (Figure 9).

Table 3. Curtailment rate by day-ahead price band, market-based, pooled over 2022 to October 2026.

| | | Wind | | | Solar | | |
|---|---|---|---|---|---|---|---|
| Price band (EUR/MWh) | Hours | Potential (TWh) | Curtailed (TWh) | Curtailment rate | Potential (TWh) | Curtailed (TWh) | Curtailment rate |
| below -100 | 44 | 0.6 | 0.4 | 57 % | 2.5 | 0.2 | 8 % |
| -100 to -50 | 62 | 1.3 | 0.6 | 51 % | 3.1 | 0.2 | 6 % |
| -50 to -20 | 176 | 3.4 | 1.2 | 35 % | 8.6 | 0.4 | 4 % |
| -20 to -10 | 167 | 3.9 | 0.8 | 22 % | 7.3 | 0.3 | 5 % |
| -10 to 0 | 1,413 | 36.8 | 5.6 | 15 % | 46.6 | 0.9 | 2 % |
| 0 to 5 (placebo) | 1,163 | 31.4 | 1.6 | 5 % | 27.4 | -0.6 | -2 % |
| above 5 (placebo) | 39,330 | 615.9 | -7.4 | -1 % | 304.9 | -2.6 | -1 % |

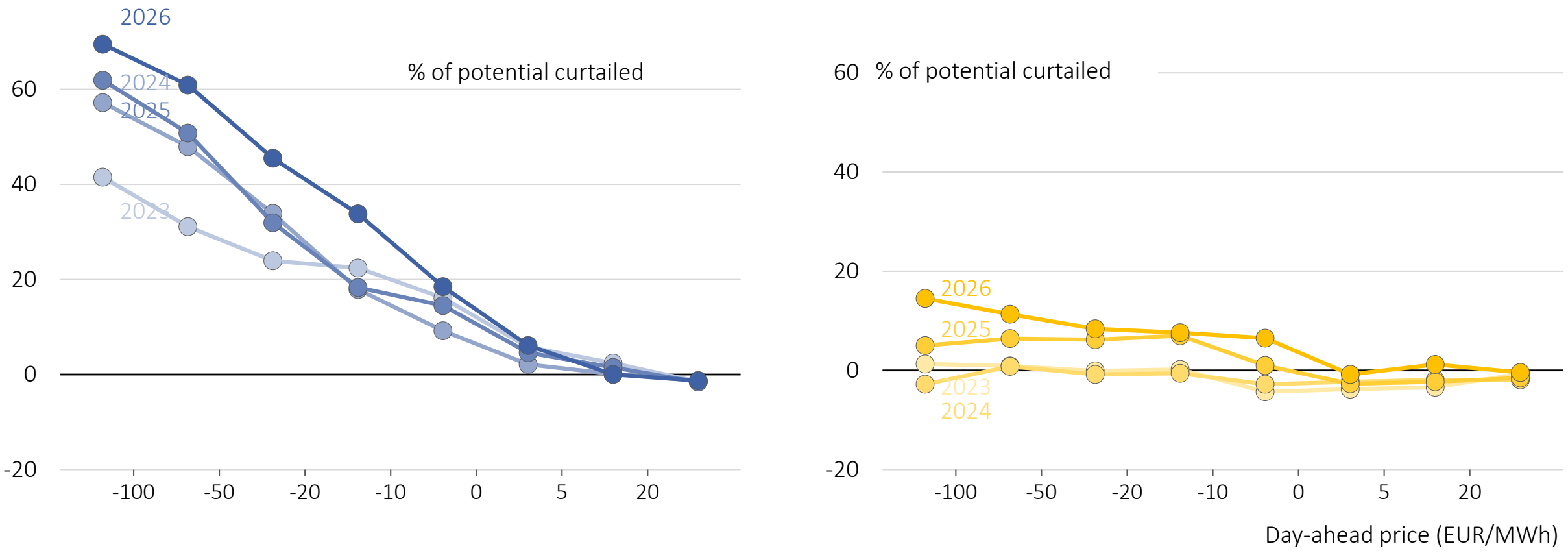

Figure 9. Curtailment rate by day-ahead price for wind (left) and solar (right).

One consequence of the persistently low solar curtailment rate despite an increase in the frequency of negative prices is that more and more solar energy is produced during negative prices (Figure 10).

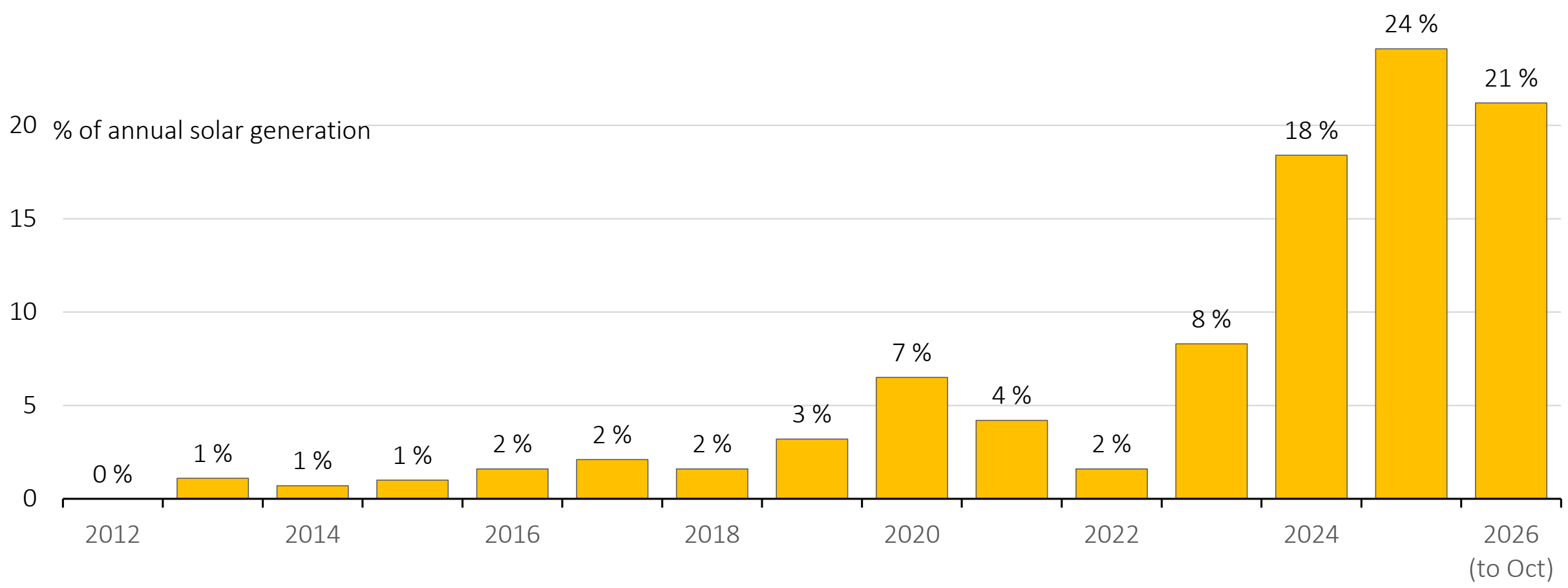

Figure 10. Share of annual solar generation produced in hours with a negative day-ahead price.

# 6. Expected versus actual curtailment

Comparing the findings of Sections 4 and 5, I now turn to the second question that this paper is trying to answer: How does expected curtailment (incentives) compare to actual curtailment (observation)? In other words, have generators acted according to incentives?

## 6.1. Methods and data

This analysis is limited to 2022 through October 2026, because no significant market-based curtailment was observed before. For every hour of that period I split the wind and solar fleet into three groups: those that have an incentive to produce and do so, those that have an incentive to curtail and do so, and those that have an incentive to curtail but keep producing nevertheless. The first two groups behave rationally, the last does not (or has other reasons to ignore incentives, such as unobserved costs).

## 6.2. Curtailment today

Across all negative-price hours of 2025, the last full year in the sample, 64 % of wind generation had an incentive to curtail (exposure rate). Of that, 27 % actually did curtail (response rate), so 17 % of the generation potential was curtailed (curtailment rate). For solar energy the numbers are even lower: only 3 % of the generation potential was curtailed during negative prices (Table 4).

Table 4. Incentive and observed behavior during the 573 negative-price hours of 2025.

| | Share of potential that has incentive to curtail (exposure rate) | Share of the exposed potential that curtails (response rate) | Share of potential that curtails (curtailment rate), exposure rate times response rate |
|---|---|---|---|
| Wind | 64 % | 27 % | 17 % |
| Solar | 28 % | 9 % | 3 % |

Both incentives and behavior depend on the hourly electricity price. If prices become more negative, two things happen: a larger fraction of generators has an incentive to curtail (by construction of the market premium), and a larger fraction of the generators that should curtail actually do so (presumably because the incentive becomes stronger). But even when prices were deeply negative (below -50 EUR/MWh), only half of wind generation and 6 % of solar generation was curtailed (Figure 11).

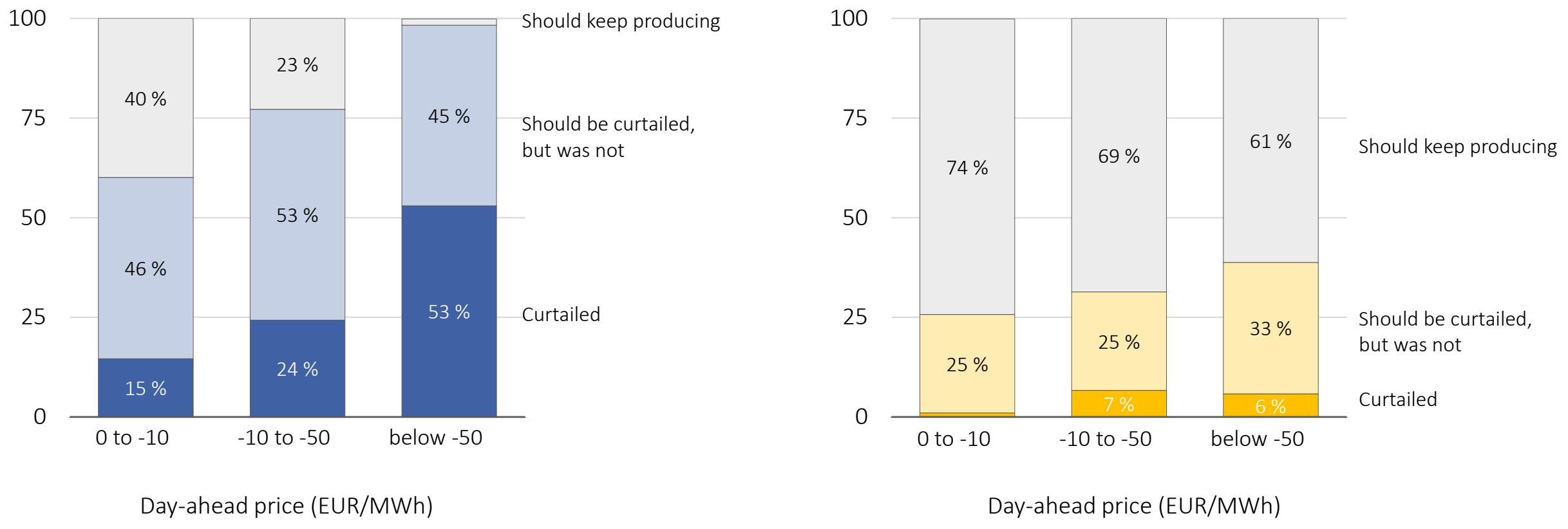

Figure 11. Incentives vs. actual curtailment by day-ahead price for wind (left) and solar (right). 2025.

## 6.3. Curtailment over time

Over the years, the exposure of wind and solar generators to negative prices has changed, because of policy reforms, swings in market prices and an expansion of installed capacity. Also their responsiveness to those prices has changed (Table 5). Figure 12 shows the exposed volume and the volume actually curtailed. For wind the response rate went from 24 % in 2022 to 37 % in 2026 without a trend in between; for solar it is not measurably different from zero before 2025 and reaches 26 % in 2026.

Table 5. Incentive and observed behavior in the negative-price hours of each year.

| | | Wind | | | Solar | | |
|---|---|---|---|---|---|---|---|
| Year | Negative hours | Exposure rate | Curtailment rate | Response rate | Exposure rate | Curtailment rate | Response rate |
| 2022 | 70 | 92 % | 22 % | 24 % | 28 % | 0 % | -1 % |
| 2023 | 300 | 56 % | 18 % | 32 % | 26 % | -3 % | -10 % |
| 2024 | 457 | 57 % | 15 % | 26 % | 23 % | -2 % | -8 % |
| **2025** | 573 | 64 % | 17 % | 27 % | 28 % | 3 % | 9 % |
| **2026 (to Oct)** | 462 | 68 % | 25 % | 37 % | 29 % | 7 % | 26 % |

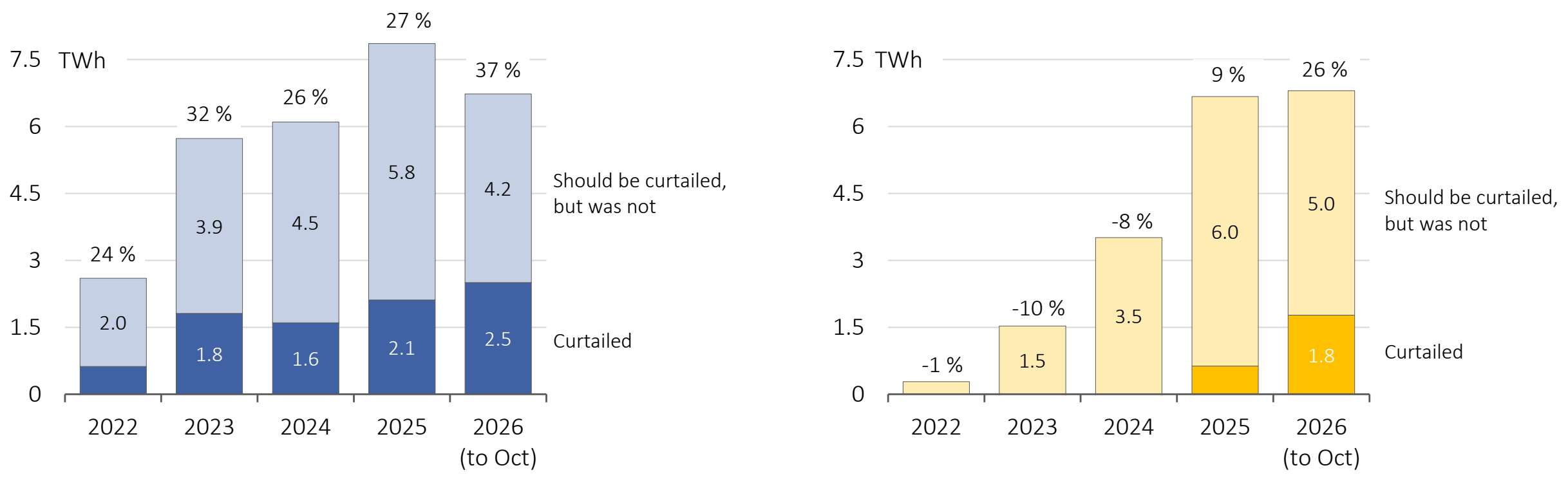

Figure 12. Incentives vs. actual curtailment over time for wind (left) and solar (right). Percentages give the ratio of the two (response rate).

## 6.4. Curtailment by strength of incentive

From the support payments I determine the profit margin of each cohort of wind and solar capacity in each hour; a negative margin is the financial incentive to curtail. Because curtailment is not observed cohort by cohort, one common response function is fitted across the classes. The two classes with a positive margin are a placebo: here, no curtailment is expected. The response rate increases with the strength of the incentive (Figure 13): the more money is at stake, the more generators stop. Even where more than EUR 50 is at stake for each MWh produced, three quarters of exposed solar generation keeps producing.

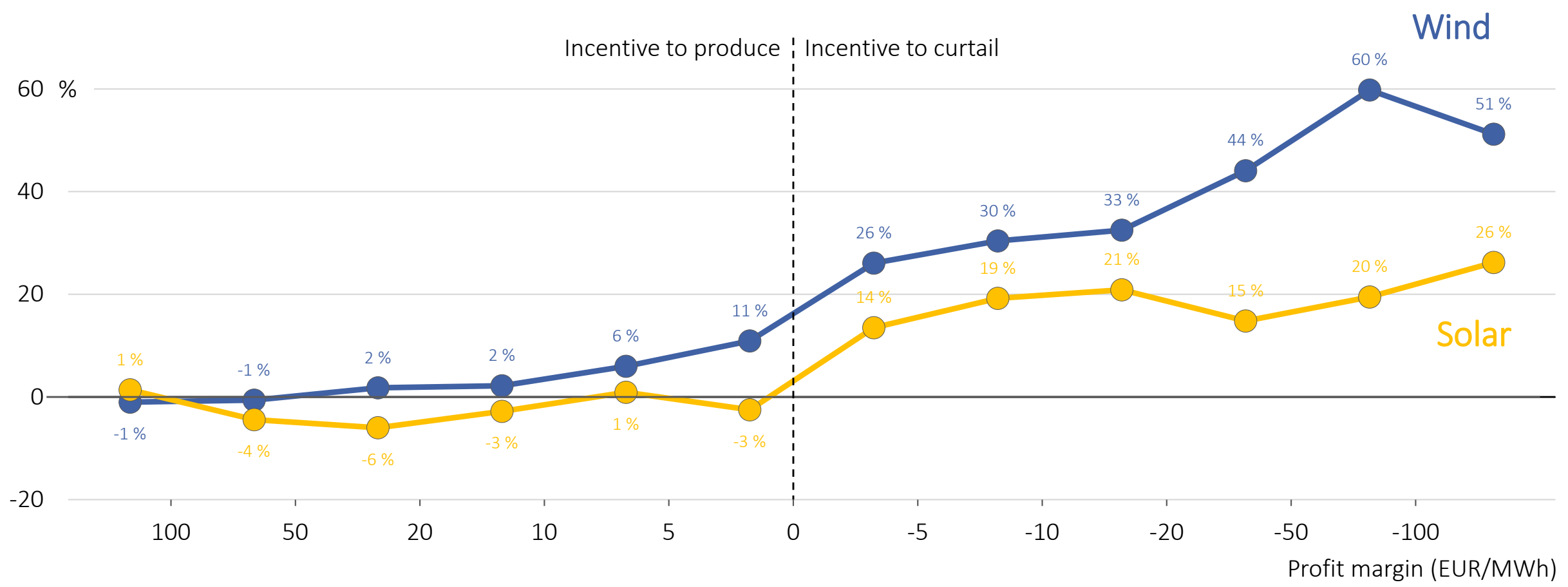

Figure 13. Share of the exposed generation potential that is curtailed, by the profit margin from producing (day-ahead price minus curtailment threshold), Germany, 2022 to October 2026.

# 7. Implications and outlook

This section discusses the implications of these findings for the present and the future.

## 7.1. Power system

I call wind and solar generation that keeps producing at deeply negative prices (below -50 EUR/MWh) “stiff generation”. It keeps producing either because it has no incentive to curtail at that price or because it does not respond to the incentive. Despite policy reforms, stiff generation in these hours has increased in recent years, reaching 48 GW in 2026 against an average load of 43 GW in the same hours. In 2026 stiff generation exceeded load in 33 of the 34 hours below -50 EUR/MWh, in 2025 in 25 of 30 (Figure 14).

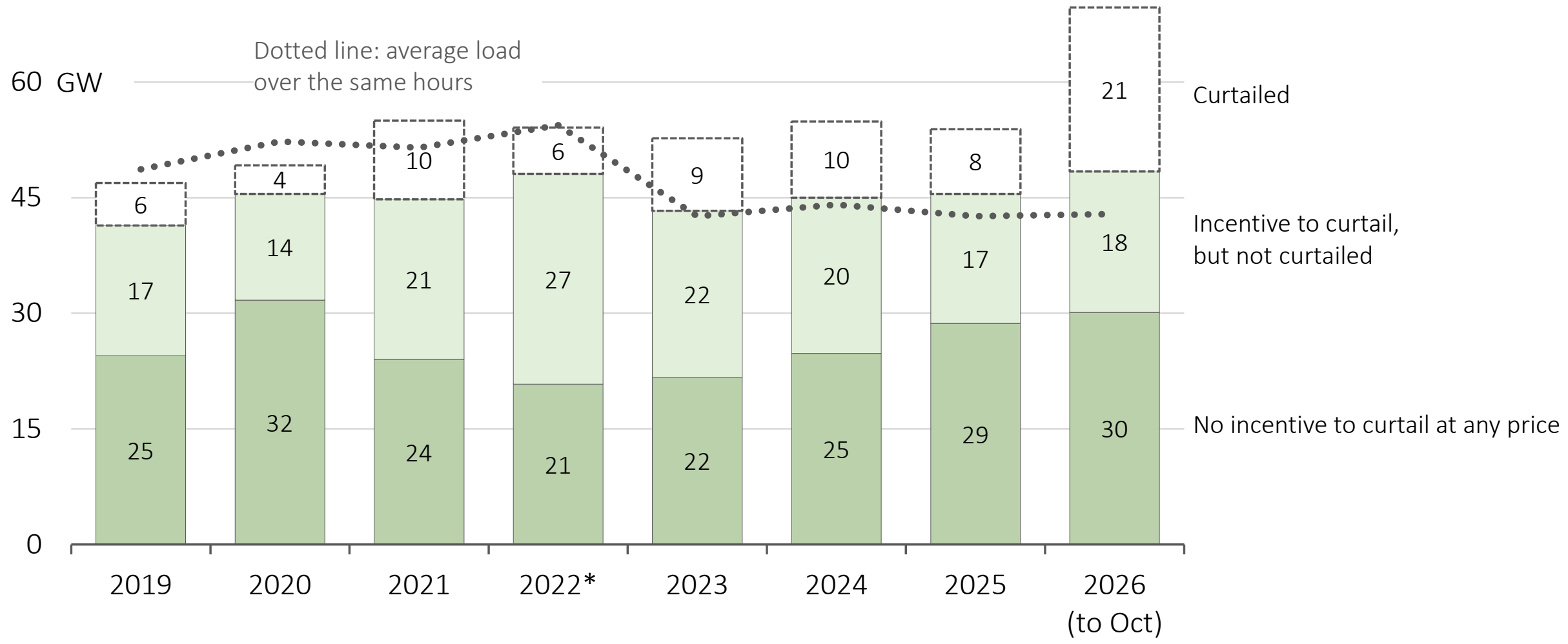

Figure 14. Stiff generation vs. load. Wind and solar generation that stays on the grid in hours with a day-ahead price below -50 EUR/MWh, 2019 to October 2026, compared to the average load over the same hours. For 2022, which had no such hour, the five hours with the lowest prices of that year are shown instead (-19 to -11 EUR/MWh); the high price level of that year implies a strong curtailment incentive already at moderately negative prices.

## 7.2. Public finance

The financial loss from selling at negative day-ahead prices is borne partly by the owners, partly by the public. I distinguish four channels (Table 6). First, the owner of a plant that keeps producing despite an incentive to curtail loses the profit margin, which is negative in these hours. Second, the public pays the support that a curtailed plant would have forgone. These two channels add up to the absolute value of the price times the volume. Third, under a feed-in tariff the support scheme bears the whole loss: the plant is paid its tariff and the EEG account sells the energy at the negative price. Fourth, generation at negative prices lowers the market value that sets the premium, because the market value is computed on the realized feed-in of the technology. The premium then rises for every supported MWh of that month, including those of plants that did curtail. Over 2022 to October 2026 the four channels add up to about EUR 2.7 billion, of which the owners bear 0.5 billion and the public 2.2 billion. The lower market value is the largest single channel with EUR 1.7 billion. The solar fleet on the feed-in tariff adds a further 0.4 billion. The negative-price provisions affect neither of these two channels. Together they are more than thirty times the support that those provisions put at stake.

Table 6. The costs of non-curtailment, EUR million.

| | Wind | | | | Solar | | | |
|---|---|---|---|---|---|---|---|---|
| Year | Owner | Support at stake | EEG account | Market value | Owner | Support at stake | EEG account | Market value |
| **2019** | 23 | 18 | 4 | 85 | 2 | 1 | 26 | 18 |
| **2020** | 21 | 16 | 5 | 69 | 9 | 2 | 49 | 36 |
| **2021** | 22 | 13 | 2 | 98 | 4 | 1 | 27 | 30 |
| **2022** | 4 | 0 | 0 | 4 | 1 | 0 | 2 | 10 |
| **2023** | 57 | 5 | 2 | 218 | 38 | 4 | 53 | 74 |
| **2024** | 44 | 8 | 2 | 247 | 61 | 9 | 82 | 219 |
| **2025** | 38 | 5 | 1 | 251 | 91 | 11 | 109 | 263 |
| **2026 (to Oct)** | 43 | 4 | 2 | 154 | 117 | 13 | 157 | 258 |
| **2022–2026** | **186** | **21** | **7** | **874** | **308** | **36** | **403** | **824** |

# 8. Conclusions and policy implications

Operating power systems with wind and solar capacity that greatly exceeds even peak demand is becoming the new normal. To avoid negative prices but also to ensure physical stability of the power grid, it is important that renewable generation is curtailed in such moments. While turning off wind turbines and solar panels is technically easier than ramping down large power stations, this paper shows for the case of Germany that this is nevertheless often not done.

The findings point to two problems: too much generation without an incentive to curtail, and too few generators responding to the incentive where it exists.

In 2025, during periods of deeply negative prices (below -50 EUR/MWh), almost all wind generators had an incentive to stop production, but only half did. In contrast, nearly two thirds of solar energy had no incentive to curtail, mostly because it received feed-in tariffs that shielded it from wholesale prices. Of those that were exposed to prices, just one in seven actually cut production.

This is no longer only a problem from a public budget perspective, but increasingly a system operations concern. In a future where wind and solar energy supply the bulk of electricity, installed capacity will be larger than peak load and its potential output will exceed electricity demand in many hours of the year. Such a system can only work if supply adjusts to demand, i.e., if wind and solar generators curtail excess energy. Germany has reached a point where this is becoming an acute problem.

# Acknowledgements

I thank Ingmar Schlecht and Jonathan Mühlenpfordt for many discussions over the past years.

# Competing interest

The author is managing director of Neon Neue Energieökonomik, a consultancy advising clients in the electricity sector. This paper did not receive any funding and was not commissioned by any client; no client has seen its results prior to publication.

# Funding

This research did not receive any specific grant from funding agencies in the public, commercial, or not-for-profit sectors.

# Data availability

The data sources are listed in Table A4. Prices, generation, capacity, market values and redispatch measures are publicly available from the transmission system operators, the ENTSO-E Transparency Platform, SMARD, AGEE-Stat and Open Power System Data. The ERA5 reanalysis is available from the Copernicus Climate Data Store. The intraday index and the day-ahead prices before 2015 are power exchange data. The derived series, in particular the hourly generation potential and the cohort-level curtailment thresholds, and the code are available from the author on request.

# CRediT contribution statement

Lion Hirth: Conceptualization, Methodology, Software, Formal analysis, Investigation, Data curation, Writing – original draft, Writing – review & editing, Visualization.

# Declaration of generative AI and AI-assisted technologies in the manuscript preparation process

During the preparation of this work, the author used Claude (Anthropic) for writing and debugging analysis code, for producing figures, and for language editing. The author reviewed and edited the output as needed and takes full responsibility for the content of the published article.

### Legal sources

# Appendix

## A1. Derivation of the curtailment threshold

Consider a price-taking plant that can be curtailed continuously and faces the day-ahead price $p$ in interval $t$. Variable operating cost is small and normalized to zero. Let the payments that accrue per MWh fed in, and are forgone if the plant does not produce, be the support at stake. Profit is linear in output,

$$\pi_{it} = (p_t + \sigma_{it} - c_i)q_{it} \tag{A1}$$

so the plant either produces at its full potential or not at all. The curtailment threshold is the price at which the bracket changes sign,

$$p_{it}{}^{*} = c_i - \sigma_{it} = -\sigma_{it} \tag{A2}$$

The plant curtails below this price. Three properties follow. First, the curtailment threshold depends on the contract and not on the plant. Identical turbines in one wind farm have different curtailment thresholds if they belong to different cohorts. Second, the curtailment threshold is independent of output, so the merit order with curtailment is a step function in installed capacity. Third, it is defined on the day-ahead price, to which the statutory provisions refer. Appendix A5 tests incentives from intraday and imbalance prices.

The support at stake follows from the settlement formula of the instrument (Table 2). Under a feed-in tariff no price triggers curtailment. I report this capacity in a separate category and do not place it at the price floor of the exchange. Under a fixed premium the support at stake is the premium. Under a one-sided CfD it is the gap between the strike price and the capture price of the settlement period, floored at zero, which the plant takes as given,

$$\sigma_{c,m} = \max\{k_c - r_m, 0\}, \qquad p_{c,m}{}^{*} = -\max\{k_c - r_m, 0\} \tag{A3}$$

where $m$ runs over settlement periods. A cohort whose capture price has risen above its strike price receives no premium and has a curtailment threshold of zero, like a merchant plant. This applied to most German cohorts in 2022. Under an hourly reference no price triggers curtailment. With certificates the curtailment threshold is minus the certificate revenue per MWh. A two-sided CfD against a period average has no floor. The payment turns into a charge once the capture price exceeds the strike price, so the curtailment threshold becomes positive,

$$p_m{}^{*} = r_m - k \tag{A4}$$

which Appendix A3 quantifies for the German revenue skim. Support that is not paid per MWh leaves the curtailment threshold at zero. Under a fixed-price pay-as-produced power purchase agreement the offtaker bears the price risk. Whether the plant stops then depends on who holds the dispatch right. For self-consumed energy the relevant price is the avoided retail tariff.

A negative-price provision withholds the payment once the day-ahead price has been negative for a given number of consecutive intervals, which differs between cohorts. The support at stake then depends on the length of the negative-price episode,

$$\sigma_{it} = \sigma_i \mathbb{1}\{L_t < N_i\}, \qquad p_{it}{}^{*} = -\sigma_i \mathbb{1}\{L_t < N_i\} \tag{A5}$$

so the curtailment threshold is zero inside an episode that is long enough and minus the full support outside it. The required length ranges from one interval to six hours, so the merit order depends on episode length as well as on price. I report the two extreme cases: an isolated negative interval, in which no suspension rule applies, and an episode of at least six consecutive negative hours, in which every suspension rule applies. Any observed episode lies between them. Rules that suspend payments from the first negative interval apply in both cases.

Where the withheld payment is returned by extending the support period, as under section 51a EEG, the plant receives it at the end of the remaining support period. With a compensation factor $\varphi$, a remaining support period $T$ and a continuous discount rate $\rho$, the value at stake is the difference between a payment today and its discounted equivalent,

$$p^{*} = -\sigma(1 - \varphi e^{-\rho T}) \tag{A6}$$

In the calculation I set the bracket to zero, that is, I treat the extension as full compensation. This is exact only for full compensation at a zero discount rate. With a remaining support period of fifteen years and a discount rate of 6 %, the curtailment threshold would be 0.59 of the support at stake for wind and 0.80 for solar, where only half of the intervals are compensated. The simplification puts the curtailment thresholds of these cohorts closer to zero than they are and thus overstates the capacity that has an incentive to curtail. I keep it because the remaining support period and the discount rate differ between plants and are not observed.

Where the reference is the realized capture price of the whole fleet, a plant that curtails raises that reference and lowers the premium of every plant in the fleet, including its own. For fleet output $Q$ in the settlement period,

$$\frac{\partial r_m}{\partial q_{it}} = \frac{p_t - r_m}{Q_m} \tag{A7}$$

which is negative at negative prices and of the order of one over the number of plants. A single plant ignores this effect, so the collective incentive to curtail exceeds the private one. Section 7.2 quantifies the reverse effect: generation at negative prices lowers the capture price and raises the premium of every supported plant.

For each cohort I use the installed capacity at the end of the year and compute the curtailment threshold for every settlement period. The merit order with curtailment is the capacity whose curtailment threshold is at or above a given price,

$$M_y(p) = \sum_c K_{c,y} \sum_m \omega_{m,y} \mathbb{1}\{p_{c,m,y}{}^* \geq p\}, \qquad \omega_{m,y} = \frac{h_{m,y}}{\sum_m h_{m,y}} \tag{A8}$$

where the weight of a settlement period is its share in the negative hours of the year and, in the episode variant, in the hours inside episodes of at least six consecutive negative hours (Table A2). I weight by negative hours and not by calendar time, because the incentive only matters when the price is negative. A cohort can therefore fall into two bands within one year. Installed capacity minus the expression above is the short-run supply of the fleet at that price. This is how the figures of Section 4 are to be read.

I group curtailment thresholds in bands of 10 EUR/MWh down to -150 EUR/MWh, with a residual band below and a separate category for capacity with no incentive at any price. Appendix A6 uses the coarser bands defined in Table A8. Capacity in band A has a curtailment threshold of zero and thus an incentive to curtail at any negative price. Bands are cumulative, so a plant in band C also has an incentive to curtail at -60 EUR/MWh.

# A2. German support schemes: provisions and cohort parameters

Germany has supported renewable energy since 2000 under the Erneuerbare-Energien-Gesetz (EEG). The EEG grants a payment per kWh fed in for twenty calendar years plus the year of commissioning. The support level (anzulegender Wert) is set by statute or, for auctioned capacity, by the awarded bid. Until the EEG 2012 it was paid as a fixed feed-in tariff. Since then plants may sell their output themselves and receive a one-sided CfD, the market premium. Above certain size thresholds they must do so. The market premium equals the support level minus the capture price of the technology, floored at zero, so the support level is the strike price of the one-sided CfD. Anlage 1 EEG defines the capture price as the feed-in-weighted average day-ahead price of the technology over the calendar month, and over the year for plants commissioned or awarded from 2023. Direct marketing became mandatory above 500 kW in 2014 and above 100 kW in 2016. Since 2017 auctions set the support level for onshore wind and for larger ground-mounted solar.

Under section 51 EEG the payment is withheld for the whole of a negative-price episode once the day-ahead price has been negative for a given number of consecutive hours. The EEG 2014 introduced this suspension rule with six hours for plants commissioned from 2016. The EEG 2021 shortened it to four hours for plants commissioned in 2021 and 2022. The EEG 2023 shortened it further in steps for the following cohort. Table A1 lists the required number of hours and the size thresholds per cohort. Plants commissioned before 2016 are not covered.

For new plants, the Solarspitzengesetz of 25 February 2025 replaced these steps by a rule that applies from the first negative interval: no payment is made in any quarter-hour with a negative day-ahead price. The rule applies from 2 kW and, below 100 kW, only once a smart meter is installed. The withheld intervals are added to the support period under section 51a EEG, at half weight for solar. Since 1 October 2025 the day-ahead auction clears quarter-hourly, so the rule and the settlement interval coincide. In addition, new small solar without a smart meter may feed in no more than 60 % of installed capacity. This reduces feed-in but gives the plant no price signal.

Three groups have no incentive to curtail at any hourly price. The first is rooftop solar below 100 kW. It receives the feed-in tariff or, after the twenty-year support period, a payment based on the annual market value. Most of it cannot be controlled remotely. The second is self-consumed generation, for which the avoided retail tariff is the relevant price. The third is plants under fixed-price pay-as-produced power purchase agreements, where the offtaker bears the price risk. Merchant plants and plants whose support period has ended face the hourly price and have a curtailment threshold of zero. I treat plants on the transitional post-support remuneration, which is based on the annual market value, in the same way. This is a simplification. The direction of the resulting error is unclear.

Support for offshore wind developed differently. The wind farms of 2010 and 2011 received the EEG 2009 tariff including the acceleration bonus. The wind farms of the compression model (Stauchungsmodell) receive a high rate for eight years and the base rate thereafter. Their curtailment threshold is therefore far below zero during the first eight years and zero afterwards. Auctions resulted in zero bids from 2017, so the wind farms commissioned from 2024 are either merchant or have a strike price from the 2017 and 2018 auction rounds. I count two wind farms as merchant because their strike price is not published.

Between December 2022 and June 2023 the Strompreisbremsegesetz skimmed 90 % of the revenues of plants above 1 MW that exceeded a ceiling. The ceiling was the support level plus a safety margin of 3 ct/kWh and 6 % of the monthly capture price. Revenues were not the plant's own hourly revenues but were calculated from the monthly capture price. The skim was therefore a two-sided settlement against a period average as in Equation A4. Appendix A3 quantifies it. Regulation (EU) 2024/1747 makes two-sided contracts for difference the default for new support. With an hourly reference such a contract has no curtailment threshold. With a period reference the curtailment threshold can become positive.

I group the fleet into 30 cohorts, defined by technology, commissioning year, support instrument, size class and marketing route (Table A1). Of these, 28 have capacity in 2025. Sections 4 to 6 use the capacity-weighted strike price of each cohort. Within an onshore wind cohort the strike price also differs by resource quality, because the reference-yield model grants poorer sites a higher support level. As a sensitivity, I split the cohorts by resource quality, which gives 84 sub-cohorts (Appendix A3). Each of the five onshore cohorts from 2016 is split into twelve quality classes. The 2003 to 2011 cohort in the market premium is split into its initial and its base tariff phase, which is how the support level depended on resource quality before 2017. Cohort capacities are my own estimates, calibrated so that they sum to the official national total for every technology and year. I base them on the capacity in direct marketing reported by the transmission system operators, the size distribution of the plant register and the auction results for strike prices after 2017. The last column of Table A1 gives the status of each entry.

Table A1. German cohorts, support parameters and curtailment thresholds.

| Cohort | Instrument and reference | Support level | Negative-price provision | GW 2025 | p* isolated | p* 6 h | Status |
|---|---|---|---|---|---|---|---|
| Onshore: up to 2002 cohort, in market premium | One-sided CfD, monthly | 8.0 ct/kWh (blended) | None (pre-2016) | 0.00 | n/a | n/a | A |
| Onshore: up to 2002 cohort, feed-in tariff | Feed-in tariff | 8.0 ct/kWh | None | 0.00 | n/a | n/a | A |
| Onshore: 2003 to 2011 cohort, in market premium | One-sided CfD, monthly | 8.5 ct/kWh (blended) | None (pre-2016) | 9.69 | -38 to 0 | -38 to 0 | A |
| Onshore: 2003 to 2011 cohort, feed-in tariff | Feed-in tariff | 8.5 ct/kWh | None | 0.51 | none | none | A |
| Onshore: 2012 to 2015 cohort, in market premium (no suspension rule) | One-sided CfD, monthly | 8.9 ct/kWh incl. management premium | None; s. 51 applies from 2016 | 14.25 | -34 to -5 | -34 to -5 | V/A |
| Onshore: 2012 to 2015 cohort, feed-in tariff | Feed-in tariff | 8.9 ct/kWh | None | 0.75 | none | none | A |
| Onshore: 2016 to 2020 cohort, turbines from 3 MW (six-hour rule) | One-sided CfD, monthly | 7.6 ct/kWh (blended) | 6 h, s. 51 EEG 2017; turbines from 3 MW | 10.05 | -41 to 0 | 0 | V/A |
| Onshore: 2016 to 2020 cohort, turbines below 3 MW (exempt) | One-sided CfD, monthly | 7.6 ct/kWh (blended) | None (below the 3 MW threshold) | 5.65 | -41 to 0 | -41 to 0 | V/A |
| Onshore: 2021 to 2022 cohort (four-hour rule) | One-sided CfD, monthly | 7.55 ct/kWh (bid 6.16 × 1.226) | 4 h, s. 51 EEG 2021; from 500 kW | 4.20 | -41 to 0 | 0 | V/A |
| Onshore: 2023 to 24 Feb 2025 cohort (EEG 2023 schedule) | One-sided CfD, monthly | 7.22 ct/kWh (bid 5.89 × 1.226) | 4/3/2/1 h schedule, s. 51 EEG 2023; from 400 kW; modeled as 3 h | 7.80 | -37 to 0 | 0 | V/A |
| Onshore: from 25 Feb 2025 cohort (any negative quarter-hour) | One-sided CfD, monthly | 8.93 ct/kWh (bid 7.28 × 1.226) | Any negative quarter-hour; time compensation s. 51a | 3.20 | 0 | 0 | V/A |
| Onshore: Post-EEG (after 20 years) / merchant / PPA | Merchant, PPA or post-support | none at stake | n/a | 11.97 | 0 | 0 | A |
| Offshore: alpha ventus / Baltic 1 (2010 and 2011) | One-sided CfD, monthly | 15.0 ct/kWh for 12 yrs, then 3.5 | None | 0.11 | 0 | 0 | A |
| Offshore: BARD + 2014 and 2015 parks (Stauchungsmodell) | One-sided CfD, monthly | 19.4 ct/kWh for 8 yrs, then 3.9 | None (pre-2016) | 3.17 | 0 | 0 | A |
| Offshore: 2016 to 2017 parks (Stauchung; 6h rule) | One-sided CfD, monthly | 19.4 ct/kWh for 8 yrs, then 3.9 | 6 h, s. 51 | 2.13 | 0 | 0 | A |
| Offshore: 2018 to 2021 parks (Stauchung; 6h rule) | One-sided CfD, monthly | 18.9 ct/kWh for 8 yrs, then 3.9 | 6 h, s. 51 | 2.38 | -124 to -91 | 0 | A |
| Offshore: 2022 to 2024 auction parks with strike | One-sided CfD, monthly | 4.6 to 6.5 ct/kWh (auction) | 3 h to 4 h by commissioning year | 0.73 | -5 to 0 | 0 | A |
| Offshore: Zero-bid parks (merchant/PPA) | Merchant or PPA | none at stake | n/a | 0.95 | 0 | 0 | A |
| Solar: Rooftop below 100 kW, feed-in tariff, not controllable | Feed-in tariff or annual-market-value payment | various | None; not remotely controllable | 56.00 | none | none | A |
| Solar: PV from 100 kW up to 2015, in market premium (no suspension rule) | One-sided CfD, monthly | 18.0 ct/kWh (blended 2009 to 2015) | None (pre-2016) | 9.30 | -160 to -67 | -160 to -67 | A |
| Solar: PV from 100 kW up to 2015, still feed-in tariff | Feed-in tariff | 18.0 ct/kWh | None | 6.20 | none | none | A |
| Solar: PV 2016 to 2020, from 500 kW (six-hour rule) | One-sided CfD, monthly | 7.0 ct/kWh (blended) | 6 h, s. 51; from 500 kW | 3.00 | -50 to 0 | 0 | V/A |
| Solar: PV 2016 to 2020, 100 to 500 kW (exempt) | One-sided CfD, monthly | 7.0 ct/kWh (blended) | None (below 500 kW) | 3.00 | -50 to 0 | -50 to 0 | V/A |
| Solar: PV 2021 to 2022, from 500 kW (four-hour rule) | One-sided CfD, monthly | 5.8 ct/kWh | 4 h, s. 51 EEG 2021 | 3.35 | -38 to 0 | 0 | V/A |
| Solar: PV 2021 to 2022, 100 to 500 kW (exempt) | One-sided CfD, monthly | 5.8 ct/kWh | None (below 500 kW) | 2.75 | -38 to 0 | -38 to 0 | V/A |
| Solar: PV 2023 to 24 Feb 2025, from 400 kW (EEG 2023 schedule) | One-sided CfD, monthly | 6.5 ct/kWh | 4/3/2/1 h schedule; from 400 kW; modeled as 3 h | 9.72 | -45 to 0 | 0 | V/A |
| Solar: PV 2023 to 24 Feb 2025, 100 to 400 kW (exempt) | One-sided CfD, annual | 6.5 ct/kWh | None (below 400 kW) | 6.48 | -20 | -20 | V/A |
| Solar: PV from 100 kW, from 25 Feb 2025 (any negative quarter-hour) | One-sided CfD, monthly | 5.5 ct/kWh | Any negative quarter-hour; time compensation s. 51a, factor 0.5 | 6.70 | 0 | 0 | V/A |
| Solar: Merchant/PPA PV, exposed to hourly price | Merchant or financial PPA | none at stake | n/a | 9.45 | 0 | 0 | A |
| Solar: Merchant/PPA PV, fixed-price pay-as-produced (not exposed) | Pay-as-produced PPA, fixed price | none at stake | Offtaker bears the price | 1.05 | none | none | A |

# A3. Reference prices and settlement periods

The curtailment threshold of a cohort under the one-sided CfD is not fixed by policy. It equals minus the difference between the strike price and the capture price of the settlement period and therefore changes with market prices. I evaluate each cohort on the period on which it settles. For the German market premium this is the calendar month, and the year for plants commissioned or awarded from 2023 (Anlage 1 Nr. 2 EEG). Either date suffices, so auctioned capacity awarded before 2023 stays on the monthly reference whatever its commissioning year. Only the non-auctioned solar of 100 to 400 kW built from 2023 settles annually (Table A1). I do not separate solar plants of 400 kW to 1 MW, which stay on the monthly reference. This affects the share of solar capacity with a curtailment threshold of zero by less than half a percentage point. The reference period matters because solar capture prices are lowest in the months in which negative prices occur. A monthly reference therefore raises the premium at stake in the months with negative prices, while an annual reference can keep the premium of a whole cohort at zero for the year. Most of the literature applies an annual reference to cohorts that settle monthly and therefore measures a different incentive.

I do not use the official monthly market values. Instead, I reconstruct the monthly series from hourly data and scale it to the official annual value. For technology $j$ and month $m$,

$$r_{j,m} = \lambda_{j,y} \frac{\sum_{t\in m} p_t g_{j,t}}{\sum_{t\in m} g_{j,t}}, \qquad \lambda_{j,y} = r_{j,y}{}^{off} \frac{\sum_{t\in y} g_{j,t}}{\sum_{t\in y} p_t g_{j,t}} \tag{A9}$$

with hourly generation per technology and hourly day-ahead prices from Appendix A4 and the official annual market value in the scaling factor. Only the profile within the year is estimated, while the annual level is the official value. Without scaling, the reconstruction deviates from the official annual values by up to 4 % for solar and onshore wind and by up to 6 % for offshore wind. Table A2 reports the monthly series and the aggregation weights.

I compute curtailment thresholds for each month and allocate the capacity of a cohort to the band of that month, weighted as in Equation A8. The weights are concentrated in the second quarter and, in the episode variant, in the long episodes (Table A2). I therefore report the two variants separately and do not average them.

The bands assume that support cannot become a charge per MWh. This does not hold for two-sided settlement against a period average. Under the revenue skim the plant paid 90 % of the excess of the monthly capture price over its ceiling, irrespective of its own hourly revenues. Marginal revenue was therefore the market price minus this charge, so the curtailment threshold was positive,

$$p_m{}^* = 0.9(r_m - k - 30)^+ \tag{A10}$$

For the seven months in which the skim was in force, Table A3 evaluates Equation A10. In December 2022 a profit-maximizing plant should have curtailed at clearly positive prices. From March 2023 capture prices were below every ceiling, so the skim no longer affected curtailment thresholds. The bands of Section 4 exclude the skim and cover the regular support schemes only.

Five assumptions affect how capacity is assigned to bands. First, cohort capacities are my own estimates calibrated to official totals. Second, the average strike price of an onshore wind cohort ignores the dispersion within the cohort that results from the reference-yield model. Before 2017 the model varied the length of the higher initial tariff. Since 2017 it applies a correction factor to the awarded bid. To test the effect, I split the cohorts by resource quality into the 84 sub-cohorts of Appendix A2. I calibrate the split on published distributions of site quality, one for each regime, because the reference site was redefined with the EEG 2017. For the fleet before auctions this is a normal distribution with mean 83 and standard deviation 12 points (Fachagentur Wind, 2016). For the auctioned fleet it is the histogram of awarded turbines (Fachagentur Wind und Solar, 2025), which implies a mean correction factor of 1.226. The split preserves the mean, so the level of support is unchanged. It raises the share of German wind capacity with a curtailment threshold of zero in 2025 from 37 to 39 %. Solar is unchanged, because it has neither a correction factor nor a two-step tariff. The effect is small because the support level does not matter where a suspension rule applies. Third, I count two offshore wind farms as merchant, because their strike prices are not published. Fourth, I assume that 90 % of merchant and power-purchase-agreement capacity faces the hourly price. Fifth, I assume that 64 % of the 2016 to 2020 onshore cohort consists of turbines of at least 3 MW, which fall under the six-hour rule. The combined effect of these assumptions is smaller than the difference between the two episode variants.

Table A2. German monthly capture prices and aggregation weights, 2021 to 2025. Capture prices in EUR/MWh, reconstructed from hourly day-ahead prices and hourly generation and scaled to the official annual market value (Equation A9). Negative hours: hours with a negative day-ahead price; of which in episodes of at least six consecutive negative hours.

| Series | Jan | Feb | Mar | Apr | May | Jun | Jul | Aug | Sep | Oct | Nov | Dec |
|---|---|---|---|---|---|---|---|---|---|---|---|---|
| 2021 wind onshore | 45 | 42 | 34 | 43 | 41 | 61 | 67 | 70 | 114 | 107 | 136 | 159 |
| 2021 solar | 55 | 45 | 41 | 46 | 42 | 69 | 74 | 77 | 117 | 128 | 183 | 271 |
| 2021 negative hours | 0 | 9 | 27 | 22 | 38 | 9 | 11 | 11 | 0 | 7 | 0 | 5 |
| 2021 of which 6 h and more | 0 | 9 | 8 | 17 | 22 | 0 | 6 | 11 | 0 | 7 | 0 | 0 |
| 2022 wind onshore | 144 | 120 | 224 | 143 | 153 | 216 | 308 | 502 | 314 | 141 | 151 | 167 |
| 2022 solar | 179 | 119 | 208 | 146 | 152 | 190 | 262 | 400 | 318 | 129 | 154 | 250 |
| 2022 negative hours | 5 | 4 | 6 | 5 | 16 | 3 | 2 | 0 | 0 | 0 | 0 | 29 |
| 2022 of which 6 h and more | 1 | 0 | 6 | 0 | 7 | 0 | 0 | 0 | 0 | 0 | 0 | 29 |
| 2023 wind onshore | 85 | 101 | 82 | 86 | 78 | 88 | 59 | 65 | 84 | 67 | 73 | 44 |
| 2023 solar | 123 | 123 | 89 | 80 | 54 | 71 | 52 | 75 | 74 | 68 | 85 | 66 |
| 2023 negative hours | 13 | 0 | 9 | 11 | 33 | 20 | 56 | 23 | 22 | 38 | 3 | 72 |
| 2023 of which 6 h and more | 13 | 0 | 7 | 0 | 22 | 6 | 48 | 23 | 6 | 16 | 0 | 61 |
| 2024 wind onshore | 64 | 53 | 55 | 49 | 59 | 58 | 53 | 65 | 64 | 68 | 89 | 72 |
| 2024 solar | 75 | 59 | 50 | 39 | 33 | 44 | 36 | 43 | 45 | 67 | 101 | 112 |
| 2024 negative hours | 16 | 4 | 12 | 50 | 78 | 64 | 81 | 68 | 40 | 25 | 11 | 8 |
| 2024 of which 6 h and more | 6 | 0 | 0 | 39 | 45 | 49 | 54 | 52 | 6 | 16 | 6 | 7 |
| 2025 wind onshore | 84 | 112 | 77 | 75 | 63 | 55 | 80 | 68 | 64 | 57 | 86 | 81 |
| 2025 solar | 113 | 108 | 50 | 31 | 21 | 20 | 59 | 38 | 43 | 70 | 90 | 92 |
| 2025 negative hours | 14 | 0 | 30 | 75 | 129 | 141 | 12 | 64 | 60 | 48 | 0 | 0 |
| 2025 of which 6 h and more | 14 | 0 | 16 | 32 | 94 | 104 | 6 | 43 | 36 | 45 | 0 | 0 |

Table A3. Curtailment thresholds implied by the revenue skim of December 2022 to June 2023, EUR/MWh (Equation A10). Positive values mean that a profit-maximizing plant should curtail at that positive price. Ceiling = support level (anzulegender Wert) plus 3 ct/kWh plus 6 % of the monthly capture price (section 16 StromPBG); the column reports the fixed part. Monthly capture prices as in Table A2. Plants of 1 MW and below were exempt.

| Cohort | Ceiling | Dec 2022 | Jan 2023 | Feb 2023 | Mar 2023 | Apr 2023 | May 2023 | Jun 2023 |
|---|---|---|---|---|---|---|---|---|
| Wind onshore, 2003–2011 cohort | 115 | 38 | 0 | 0 | 0 | 0 | 0 | 0 |
| Wind onshore, 2012–2015 cohort | 119 | 34 | 0 | 0 | 0 | 0 | 0 | 0 |
| Wind onshore, 2016–2020 cohort | 106 | 46 | 0 | 0 | 0 | 0 | 0 | 0 |
| Wind onshore, 2021–2022 cohort | 95 | 56 | 0 | 0 | 0 | 0 | 0 | 0 |
| Solar, plants of 2009 to 2015 | 210 | 22 | 0 | 0 | 0 | 0 | 0 | 0 |
| Solar, plants of 2016–2020 | 100 | 122 | 14 | 14 | 0 | 0 | 0 | 0 |
| Solar, plants of 2021–2022 | 88 | 132 | 25 | 25 | 0 | 0 | 0 | 0 |

# A4. Data: sources and construction

Table A4 lists all data series with source, coverage, resolution and treatment. The following paragraphs describe how I process them.

Day-ahead prices are the exchange prices of the German bidding zone, which included Austria until September 2018. From 2015 I take them from the ENTSO-E Transparency Platform, for earlier years from the Phelix series of the power exchange. The two sources are identical in the overlapping years, except for a few dozen hours in 2019 and 2024. I average the quarter-hourly prices introduced in October 2025 to hourly values. All series are in UTC.

For intraday prices I use the quarter-hourly ID3 index, aggregated to hourly means and to the number of negative quarter-hours per hour. I use it only in Appendix A5, because the statutory provisions refer to the day-ahead price.

Metered generation is from SMARD, the data platform of the Federal Network Agency, from 2015 and from Open Power System Data (2019) for earlier years. The latter does not separate onshore and offshore wind. Both sources report feed-in at the transmission level. I scale the hourly series year by year to gross generation, which is consistent with gross installed capacity. The difference between gross generation and feed-in is constant over time for wind. For solar it increases as self-consumption grows. Hirth (2026) describes the procedure.

Installed capacity is gross capacity at the end of the year from AGEE-Stat, interpolated within the year and extended for 2026 with monthly additions from the plant register. I do not use the capacity reported on the transparency platform, because it starts only in 2018, its entry for the current year is a placeholder and it is below the official solar total. With the AGEE-Stat series, the bias in a leave-one-year-out test varies half as much (Hirth, 2026).

Grid-related curtailment is from the redispatch records of the transmission system operators, which are available per measure from October 2021. The records do not distinguish onshore from offshore wind, so I assign each measure by the name of the asset. A split in proportion to generation would misallocate most of the offshore curtailment. Before October 2021 only annual totals of feed-in management are available. I allocate them to hours in proportion to generation. Hirth (2026) documents the procedure.

Table A4. Data sources and treatment.

| Series | Source | Coverage | Resolution | Treatment |
|---|---|---|---|---|
| Day-ahead price, German market area | ENTSO-E Transparency Platform | 2015 to 2026 | hourly; quarter-hourly from Oct 2025 | DE-AT-LU to Sep 2018, DE-LU thereafter; quarter-hours averaged |
| Day-ahead price, German market area | Exchange Phelix day-ahead series | 2001 to 2014 | hourly | converted from local time; cross-checked against the platform for 2015 to 2024 |
| Intraday index ID3 | Power exchange | Jul 2020 to Dec 2025 | quarter-hourly | hourly mean, hourly minimum and count of negative quarter-hours |
| Imbalance price (reBAP) | Transmission system operators | 2015 to 2026 | quarter-hourly | hourly mean and minimum |
| Metered generation by technology | SMARD (Federal Network Agency) | 2015 to 2026 | hourly | grid feed-in at transmission level |
| Metered generation, wind and solar | Open Power System Data, release of 5 June 2019 | 2010 to 2014 | hourly | wind from 2010, solar from 2012; no offshore split |
| Gross generation by technology | AGEE-Stat | 2001 to 2025 | annual | used to scale the hourly series to a gross basis |
| Installed capacity | AGEE-Stat; plant register for monthly additions | 1990 to 2026 | annual, monthly from 2025 | linear interpolation between year-end anchors |
| Annual market values by technology | Transmission system operators | 2012 to 2025 | annual | level anchor of the monthly reconstruction |
| Redispatch measures | Transmission system operators | Oct 2021 to 2026 | per measure | technology split by asset name; aggregated to hourly energy |
| Feed-in management volumes | Monitoring reports of the federal network agency | 2009 to 2021 | annual | allocated in proportion to generation; technology split from 2013 |
| Wind speed at 100 m, surface irradiance | ERA5 reanalysis (Hersbach et al., 2020) | 2001 to 2026 | hourly, 0.25 degrees | 12 onshore and 5 offshore nodes; instantaneous wind averaged to interval means |

# A5. Robustness of the empirical estimates

Market-based curtailment in hour $t$ is the residual

$$C_{j,t}{}^{mkt} = P_{j,t} - G_{j,t} - C_{j,t}{}^{grid} \tag{A11}$$

with the generation potential from Section 5.1, metered generation scaled to gross generation and recorded grid-related curtailment. I do not truncate the residual at zero, so model error of either sign remains visible. The size of the residuals in hours without an incentive to curtail indicates the precision of the method. The annual technology coefficient of the potential model (Hirth, 2026) captures any model error that is the same in all hours. Model error that is specific to hours with low prices is attributed to market-based curtailment. This is the main limitation of the estimate. Table A5 reports the annual volumes.

The years 2010 to 2014 serve as a placebo. In 2010 and 2011 all wind and solar capacity received the feed-in tariff. From 2012 wind capacity moved to the market premium, about half of it in 2012 and more than 80 % from 2013. The curtailment thresholds of this capacity were mostly below -25 EUR/MWh (Table A8), so the day-ahead price rarely fell below them. In the negative-price hours of 2013 and 2014, 4 % of the wind potential had an incentive to curtail. In 2012 it was 19 %. Solar had no incentive to curtail in any of these hours. The measured difference in these years is therefore mostly model error. Over the negative-price hours of those years the residual is within ±0.1 TWh a year for wind and within ±0.01 TWh for solar. This is well below the volumes from 2019 onwards. Over all hours of the year the residual is larger, between -0.7 and 0.0 TWh a year for onshore wind and between -0.2 and +0.1 TWh for solar. This is model error that the annual technology coefficient does not capture. I nevertheless report wind curtailment only from 2022, because until October 2021 grid-related curtailment is available only as an annual total, which I allocate to hours in proportion to generation. For solar, the annual residual guided the specification. With a linear irradiance factor the model shows curtailment in the placebo years. A correction of the solar series in twenty irradiance bins removes most of it (Hirth, 2026). The remainder limits the precision of the solar estimate. For this reason I make no statement on solar curtailment before 2025.

The statutory provisions refer to the day-ahead price, but a direct marketer can also respond to the intraday price and is settled at the imbalance price, which is known only after delivery. Table A6 distinguishes the three prices. Hours with a negative imbalance price but a non-negative day-ahead price are frequent and show no curtailment beyond model error. Hours with a negative intraday price but a non-negative day-ahead price do show curtailment. In a regression on the hours with a negative day-ahead price, the day-ahead and the intraday price have almost the same coefficient. Adding the intraday price improves the fit, while adding the imbalance price does not. I conclude that generators respond to intraday prices, while a response to imbalance prices cannot be measured. The response appears to double in hours with a negative imbalance price, but this reflects the intraday price.

Within the hours with a negative day-ahead price, the curtailed share of the wind potential is higher at lower prices, at higher wind potential and in later years. It increases almost as much across intraday price bands as across day-ahead price bands.

The difference between wind and solar does not result from comparing different hours. At the same prices in the same daylight hours, the curtailed share of wind is several times that of solar. Before 2025 solar curtailment is within the range of model error. This is in line with the merit order of Section 4. Most of the solar fleet has no incentive to curtail at any price, compared with almost none of the wind fleet. In addition, the curtailment thresholds of the supported solar cohorts are lowest in the months with the most negative hours.

Montel (2026) provides the only other estimate of market-based curtailment for Germany. Its order of magnitude and its change over time are similar to my estimate. Montel does not disclose how it estimates the generation potential or how it treats grid-related curtailment, so the comparison is indicative only. On the incentive side, the only other study is Kern et al. (2024), who distinguish only between capacity with and without an incentive.

There are five main sources of error. The first is the model of the generation potential. The placebo residual indicates its size (Table A5). The second is that the residual may contain grid-related curtailment that is not recorded. This matters before October 2021 and for offshore wind, where outages of converter stations are not in the redispatch records. The third is the scaling to gross generation, which is constant within the year although the share of self-consumption is not. The fourth is the annual totals of feed-in management before 2013, which I have not verified against the primary source. The fifth is metered generation itself. Hourly wind and solar generation is an estimate that the transmission system operators extrapolate from a sample of plants. A bias in this estimate that is specific to negative-price hours could not be distinguished from curtailment. None of the tests in this appendix can detect such a bias.

Table A5. Annual volumes, Germany. Market-based curtailment is potential minus metered generation on a gross basis minus grid-related curtailment, summed over the hours with a negative day-ahead price; the share is of the generation potential in those hours. Negative entries are model error, not curtailment. Grid-related curtailment is the annual total. 2008 and 2009 carry only the negative-hour count, because the hourly generation series used here begins in 2010. 2010 to 2014 are placebo years with almost no incentive to curtail; the market-based figures up to 2021 are upper bounds, because grid-related curtailment is allocated pro rata to generation before October 2021. 2026 to 31 October.

| Year | Wind market TWh | Wind share of potential | Wind grid TWh | Solar TWh | Solar share of potential | Negative hours |
|---|---|---|---|---|---|---|
| 2008 | | | | | | 15 |
| 2009 | | | | | | 71 |
| 2010 | -0.02 | -10.7 % | 0.13 | | | 12 |
| 2011 | -0.01 | -5.7 % | 0.42 | | | 15 |
| 2012 | +0.05 | +4.2 % | 0.38 | -0.00 | n/a | 56 |
| 2013 | -0.02 | -2.0 % | 0.48 | -0.00 | -1.1 % | 64 |
| 2014 | -0.09 | -5.7 % | 1.22 | +0.01 | +4.1 % | 64 |
| 2015 | -0.15 | -5.6 % | 4.12 | -0.04 | -11.5 % | 110 |
| 2016 | -0.27 | -9.6 % | 3.52 | -0.02 | -3.2 % | 97 |
| 2017 | +0.01 | +0.3 % | 5.29 | -0.03 | -3.2 % | 147 |
| 2018 | -0.02 | -0.4 % | 5.25 | -0.02 | -3.4 % | 132 |
| 2019 | +0.32 | +4.1 % | 6.30 | +0.01 | +0.7 % | 211 |
| 2020 | +0.79 | +7.4 % | 5.96 | -0.09 | -3.0 % | 298 |
| 2021 | +0.28 | +5.8 % | 5.50 | +0.07 | +3.3 % | 139 |
| 2022 | +0.62 | +21.9 % | 7.34 | -0.00 | -0.2 % | 70 |
| 2023 | +1.81 | +17.6 % | 9.29 | -0.14 | -2.7 % | 300 |
| 2024 | +1.60 | +15.0 % | 7.97 | -0.27 | -1.9 % | 457 |
| 2025 | +2.11 | +17.2 % | 6.57 | +0.63 | +2.6 % | 573 |
| 2026 | +2.50 | +25.1 % | 6.18 | +1.76 | +7.4 % | 462 |

Table A6. Curtailed share of wind potential by class of hour, percent. Classes are defined by the sign of the day-ahead price and of the intraday index, with the imbalance-only class from the corresponding split by imbalance price. In classes without curtailment the values are close to zero.

| Year | Day-ahead and intraday negative | Day-ahead only | Intraday only | Imbalance only | Neither |
|---|---|---|---|---|---|
| 2021 | 6.9 % | 4.6 % | -0.4 % | -1.5 % | -0.1 % |
| 2022 | 23.2 % | 20.2 % | 11.5 % | -1.1 % | -0.8 % |
| 2023 | 22.6 % | 9.5 % | 6.9 % | -2.1 % | -0.9 % |
| 2024 | 21.0 % | 7.5 % | 2.8 % | -2.1 % | -1.0 % |
| 2025 | 22.4 % | 10.8 % | 6.8 % | -1.3 % | -1.4 % |

The suspension rules affect behavior. In episodes of at least six consecutive negative hours, the curtailed share of the wind potential is at least twice the share in short episodes (Table A7). This corresponds to the two variants of the merit order in Section 4.

Table A7. Curtailed share of potential in negative-price hours by the length of the negative episode, Germany, percent. 2026 to October. Negative entries are model noise.

| Year | Wind, 1 to 3 h | Wind, 4 to 5 h | Wind, 6 h and more | Solar, 1 to 3 h | Solar, 6 h and more |
|---|---|---|---|---|---|
| 2022 | 12.9 % | 13.8 % | 26.2 % | -0.3 % | -1.7 % |
| 2023 | 9.6 % | 10.7 % | 20.1 % | -2.8 % | -2.8 % |
| 2024 | -2.1 % | 4.8 % | 21.9 % | -5.1 % | -1.0 % |
| 2025 | 0.7 % | 7.2 % | 19.8 % | -1.7 % | 4.2 % |
| 2026 (to Oct) | 14.2 % | 11.3 % | 32.6 % | 5.9 % | 8.5 % |

# A6. Additional results: capacity by threshold band

Tables A8 and A9 report the German merit orders with curtailment in numbers. They show installed capacity by band of the curtailment threshold for every year from 2006 to 2025 for an isolated negative hour. The last column shows the capacity with a curtailment threshold of zero in an episode of at least six consecutive negative hours.

Table A8. German wind capacity by curtailment threshold band, GW, isolated negative hour. Band A: curtailment threshold of zero; B: -1 to -25; C: -25 to -50; D: -50 to -100; E: below -100 EUR/MWh; F: no incentive at any price. Last column: band A in an episode of at least six consecutive negative hours.

| Year | A | B | C | D | E | F | A, 6 h and more |
|---|---|---|---|---|---|---|---|
| 2006 | 0.0 | 0.0 | 0.0 | 0.0 | 0.0 | 20.5 | 0.0 |
| 2007 | 0.0 | 0.0 | 0.0 | 0.0 | 0.0 | 22.1 | 0.0 |
| 2008 | 0.0 | 0.0 | 0.0 | 0.0 | 0.0 | 22.8 | 0.0 |
| 2009 | 0.0 | 0.0 | 0.0 | 0.0 | 0.0 | 25.8 | 0.0 |
| 2010 | 0.0 | 0.0 | 0.0 | 0.0 | 0.0 | 27.3 | 0.0 |
| 2011 | 0.0 | 0.0 | 0.0 | 0.0 | 0.0 | 28.9 | 0.0 |
| 2012 | 0.0 | 0.1 | 5.9 | 9.4 | 0.3 | 15.4 | 0.0 |
| 2013 | 0.0 | 0.1 | 4.1 | 22.7 | 0.5 | 6.7 | 0.0 |
| 2014 | 0.0 | 0.0 | 0.8 | 31.7 | 1.1 | 5.7 | 0.0 |
| 2015 | 0.0 | 0.1 | 1.2 | 36.3 | 3.3 | 4.2 | 0.0 |
| 2016 | 0.0 | 0.2 | 3.1 | 40.5 | 4.1 | 2.1 | 3.9 |
| 2017 | 0.0 | 0.1 | 7.5 | 41.2 | 5.4 | 2.0 | 8.6 |
| 2018 | 0.0 | 1.1 | 18.3 | 31.5 | 6.4 | 2.0 | 11.2 |
| 2019 | 0.0 | 0.8 | 13.8 | 37.3 | 7.5 | 2.0 | 13.3 |
| 2020 | 0.0 | 0.2 | 7.1 | 45.9 | 7.8 | 2.0 | 14.5 |
| 2021 | 10.3 | 10.5 | 26.9 | 8.1 | 7.1 | 1.7 | 25.5 |
| 2022 | 61.4 | 0.0 | 2.5 | 0.8 | 0.0 | 1.6 | 64.1 |
| 2023 | 23.4 | 22.8 | 17.1 | 0.4 | 4.2 | 1.4 | 38.1 |
| 2024 | 19.0 | 23.3 | 24.6 | 0.1 | 4.4 | 1.4 | 42.6 |
| 2025 | 30.5 | 25.2 | 18.2 | 0.1 | 2.3 | 1.3 | 49.6 |

Table A9. German solar capacity by curtailment threshold band, GW, isolated negative hour. Bands as in Table A8.

| Year | A | B | C | D | E | F | A, 6 h and more |
|---|---|---|---|---|---|---|---|
| 2006 | 0.0 | 0.0 | 0.0 | 0.0 | 0.0 | 2.9 | 0.0 |
| 2007 | 0.0 | 0.0 | 0.0 | 0.0 | 0.0 | 4.2 | 0.0 |
| 2008 | 0.0 | 0.0 | 0.0 | 0.0 | 0.0 | 6.1 | 0.0 |
| 2009 | 0.0 | 0.0 | 0.0 | 0.0 | 0.0 | 10.6 | 0.0 |
| 2010 | 0.0 | 0.0 | 0.0 | 0.0 | 0.0 | 18.0 | 0.0 |
| 2011 | 0.0 | 0.0 | 0.0 | 0.0 | 0.0 | 25.4 | 0.0 |
| 2012 | 0.0 | 0.0 | 0.0 | 0.0 | 2.6 | 30.4 | 0.0 |
| 2013 | 0.0 | 0.0 | 0.0 | 0.0 | 5.8 | 30.5 | 0.0 |
| 2014 | 0.0 | 0.0 | 0.0 | 0.0 | 7.7 | 30.6 | 0.0 |
| 2015 | 0.0 | 0.0 | 0.0 | 0.0 | 8.7 | 31.0 | 0.0 |
| 2016 | 0.0 | 0.2 | 0.4 | 0.0 | 9.5 | 31.1 | 0.3 |
| 2017 | 0.0 | 0.1 | 1.2 | 0.0 | 9.5 | 32.0 | 0.7 |
| 2018 | 0.0 | 0.6 | 2.0 | 0.0 | 9.5 | 33.8 | 1.3 |
| 2019 | 0.5 | 0.8 | 3.1 | 0.0 | 9.5 | 35.2 | 2.4 |
| 2020 | 1.1 | 0.9 | 2.8 | 2.4 | 9.5 | 37.4 | 4.1 |
| 2021 | 4.9 | 2.6 | 3.2 | 0.5 | 8.7 | 39.5 | 7.5 |
| 2022 | 21.3 | 0.0 | 2.8 | 0.5 | 0.0 | 42.8 | 23.2 |
| 2023 | 17.6 | 6.2 | 0.0 | 1.1 | 8.3 | 48.7 | 21.0 |
| 2024 | 9.7 | 15.0 | 9.6 | 0.4 | 9.0 | 57.1 | 23.1 |
| 2025 | 18.7 | 10.3 | 14.0 | 1.7 | 9.1 | 63.2 | 33.1 |